\documentclass[reqno]{amsart}
\usepackage{latexsym,amsfonts,amssymb,cite}

\numberwithin{equation}{section}

\newcommand{\mc}[1]{{\mathcal #1}}

\newcommand{\mb}[1]{{\mathbf #1}}
\newcommand{\bb}[1]{{\mathbb #1}}

\renewcommand{\epsilon}{\varepsilon}
\renewcommand{\hat}{\widehat}

\title{ 
Thermodynamic transformations of nonequilibrium states
}

\author [L. Bertini] {Lorenzo Bertini}
\address{\noindent Lorenzo Bertini \hfill\break\indent 
Dipartimento di Matematica, Universit\`a di Roma `La Sapienza' 
\hfill\break\indent 
P.le Aldo Moro 2, 00185 Roma, Italy}
\email{bertini@mat.uniroma1.it}

\author[D. Gabrielli]{Davide Gabrielli}
\address{\noindent Davide Gabrielli \hfill\break\indent 
 Dipartimento di Matematica, Universit\`a dell'Aquila
\hfill\break\indent 
67100 Coppito, L'Aquila, Italy
}
\email{gabriell@univaq.it}

\author[G. Jona-Lasinio]{Giovanni Jona-Lasinio}
\address{\noindent Giovanni Jona-Lasinio
\hfill\break\indent 
 Dipartimento di Fisica and INFN, Universit\`a di Roma La Sapienza
\hfill\break\indent 
 P.le A.\ Moro 2, 00185 Roma, Italy
}
\email{gianni.jona@roma1.infn.it}

\author[C. Landim]{Claudio Landim} 
\address{Claudio Landim
  \hfill\break\indent IMPA \hfill\break\indent Estrada Dona Castorina
  110, \hfill\break\indent
J. Botanico, 22460 Rio de Janeiro, Brazil\hfill\break\indent
  {\normalfont and} \hfill\break\indent CNRS UMR 6085, Universit\'e de
  Rouen, \hfill\break\indent Avenue de l'Universit\'e, BP.12,
  Technop\^ole du Madril\-let, \hfill\break\indent
F76801 Saint-\'Etienne-du-Rouvray, France.} 
\email{landim@impa.br}

\begin{document}

\begin{abstract}
  We consider a macroscopic system in contact with boundary reservoirs
  and/or under the action of an external field. We discuss the case in
  which the external forcing depends explicitly on time and drives the
  system from a nonequilibrium state to another one. In this case the
  amount of energy dissipated along the transformation becomes
  infinite when an unbounded time window is considered.  Following the
  general proposal by Oono and Paniconi and using results of the
  macroscopic fluctuation theory, we give a natural definition of a
  renormalized work. We then discuss its thermodynamic relevance by
  showing that it satisfies a Clausius inequality and that quasi static
  transformations minimize the renormalized work. In addition, we
  connect the renormalized work to the quasi potential describing the
  fluctuations in the stationary nonequilibrium ensemble.  The latter
  result provides a characterization of the quasi potential that does
  not involve rare fluctuations.
\end{abstract}

\keywords{Nonequilibrium stationary states, Thermodynamic
  transformations, Clausius inequality,  Large fluctuations, Relative entropy}

\maketitle
\thispagestyle{empty}

\section{Introduction}

The basic paradigm of equilibrium statistical mechanics states that in
order to obtain the typical value of macroscopic observables and their
fluctuations we do not have to solve any equation of motion and the
calculations can be performed by using the Gibbs distribution.  The
simplest nonequilibrium states one can imagine are \emph{stationary
states} of systems in contact with different reservoirs and/or under
the action of external (electric) fields. In such cases, contrary to
equilibrium, there are currents (electrical, heat, mass,...) through
the system whose macroscopic behavior is encoded in transport
coefficients like the diffusion coefficient, the conductivity or the
mobility.  In this case we cannot bypass an analysis of the dynamical
properties of the system. Indeed, the Gibbs distribution has to be
replaced by the invariant distribution for the microscopic dynamics.
The calculation of this distribution, even for very simple models, is
a most challenging task.  On the other hand, we are mainly interested
in the macroscopic behavior of ``few'' observables and this question
may be answered without the complete microscopic knowledge of the
stationary ensemble.

A main goal that we want to reach for nonequilibrium stationary states
is therefore to construct analogues of thermodynamic potentials from
which we can extract the typical macroscopic behavior of the system as
well as the asymptotic probability of fluctuations.  As it has been
shown in \cite{primo}, this program can be implemented without the
explicit knowledge of the stationary ensemble and requires as input
the macroscopic dynamical behavior of systems which can be
characterized by the transport coefficients.  This theory, now known
as \emph{macroscopic fluctuation theory}, is based on an extension of
Einstein equilibrium fluctuation theory to stationary nonequilibrium
states combined with a dynamical point of view. It has been very
powerful in studying concrete microscopic models but can be used also
as a phenomenological theory.  It has led to several new interesting
predictions \cite{cumulanti,curr,lagpt,rev,bd,dd}.

\medskip
In this paper we develop a theory of thermodynamic transformations for
nonequilibrium stationary states. We thus consider an open system in
contact with boundary reservoirs and/or under the action of an
external field and we analyze the situation in which the reservoirs
and field vary with time driving the system from a state to another
one.  
In the case in which the initial and final states are equilibrium
states, according to the standard thermodynamic theory, the
transformation is \emph{reversible} if the energy exchanged between the
system and the environment is minimal. A thermodynamic
principle asserts that reversible transformations are accomplished by a
sequence of equilibrium states and are well approximated by
\emph{quasi static transformations} in which the
variations of the environment are very slow.  
By an explicit construction of quasi static transformations, 
we show that this principle can be derived from the assumption that
during the transformation the system is a \emph{local equilibrium}
state so that the macroscopic evolution can be described by
hydrodynamic equations.  Consider now the situation in which either
the initial or the final state is a nonequilibrium state which
supports a non vanishing current. To maintain such states one needs to
dissipate a positive amount of energy per unit of time.  If we
consider a transformation between nonequilibrium stationary states,
the energy dissipated along such transformation will necessarily
include the contribution needed to maintain such states which is
infinite when an unbounded time window is considered. Following the
general proposal in \cite{op}, we thus define a \emph{renormalized
  work} performed by a given transformation that is obtained by
subtracting from the total energy exchanged the energy needed to
maintain the (nonequilibrium) stationary state. 
We prove that the renormalized work satisfies a Clausius
inequality which states that it is larger then the
variation of the equilibrium free energy functional evaluated at the
corresponding nonequilibrium profiles. As a particular case, we
recover the equilibrium Clausius inequality. 
We finally show that also for nonequilibrium states quasi static
transformations are optimal, in the sense that  they minimize the renormalized work.

The second main topic that we here discuss is the connection between
the energy exchanged in a thermodynamic transformation and the
asymptotic probability of fluctuations in the stationary ensemble.  In
the context of equilibrium statistical mechanics, the
Boltzmann-Einstein theory, see e.g.\ \cite[Eq.~(112.2)]{LL}, states
that the probability of a fluctuation for a system in contact with an
environment at temperature $T_\mathrm{env}$, pressure
$p_\mathrm{env}$, and chemical potential $\lambda_\mathrm{env}$, is
given by
\begin{equation}
\label{be}
P \asymp \exp\Big\{- {\frac {R} {\kappa T_\mathrm{env}}} \Big\} 
\end{equation}
where $\kappa$ is the Boltzmann constant and
\begin{equation}
\label{rmin}
R = \Delta U - T_\mathrm{env} \Delta S + 
p_\mathrm{env}\Delta V -\lambda_\mathrm{env} \Delta N
\end{equation}
in which $\Delta U, \Delta S$, $\Delta V$, and $\Delta N$ are the
variations of energy, entropy, volume, and number of particles of the
system.  The quantity $R$ coincides with the variation of the
\emph{availability}, see \cite[Ch.~7]{pippard}, and it has the
interpretation of maximal useful work that can be extracted from a
system in a given environment (e.g.\ a boiler of hot water in a colder
environment). Equivalently, $R$ is the minimal work to produce the
given fluctuation.  In the context of equilibrium states we identify
$R$ with the \emph{quasi potential} introduced in the macroscopic
fluctuation theory.  Furthermore, we show that $R$ is the \emph{excess
  work} (with respect to a quasi static transformation) associated to
the transformation given by the relaxation path described by the
hydrodynamic equations which brings the system from the state
associated with the fluctuation to one in equilibrium with the
environment.

For nonequilibrium stationary states the formula \eqref{rmin} cannot
be used.  By taking a dynamical point of view we can however properly
define $R$ as the excess work with respect to the minimal renormalized
work mentioned before, so that the fluctuation formula \eqref{be}
still holds.  Furthermore, we show that the same expression can be
used to compare two different nonequilibrium stationary states. More
precisely, the relative entropy of the stationary ensembles associated
to two boundary driven stochastic lattice gases with different
boundary reservoirs can be expressed in terms of $R$.

The theory of thermodynamic transformations here developed is presented
without any reference to the underlying microscopic dynamics. 
On the other hand, the transformations here considered have well
defined microscopic counterparts. In particular, for stochastic
lattice gases it is possible to give a microscopic definition of the
work exchanged between the system and the environment. This is
a fluctuating variable whose typical behavior in the hydrodynamic
scaling limit agrees with the macroscopic description. The statistics
of the fluctuations can be derived from the fluctuations of the
empirical current \cite{curr} but will not be discussed in this paper.

\subsubsection*{Outline} $~$

In Section~\ref{s:2} we formulate the basic assumptions for the
thermodynamic description of driven diffusive systems. These
assumptions are based on the notion of local equilibrium and the
validity of the local Einstein relation.

In Section~\ref{s:es} we discuss the case of equilibrium states and
show how the Clausius inequality can be deduced from the
previous assumptions. In addition, we connect the \emph{availability}
of classical thermodynamic with the relative entropy between Gibbs
states.

In Section~\ref{s:4} we analyze the case of nonequilibrium states and
their transformations. We thus introduce the \emph{renormalized work}
performed along a given transformation and connect it to the quasi
potential of the macroscopic fluctuation theory. We also show that the
relative entropy between two different nonequilibrium states can be
expressed in terms of the quasi potential. 

In Section~\ref{s:5} we consider a system with a general time
dependent forcing. We introduce the corresponding time dependent
quasi potential which takes into account the fact that the system has
a finite relaxation time and provides a fluctuation formula for each
fixed time. We then connect also the  time dependent quasi potential
to a properly defined renormalized work. 

In Sections \ref{s:6} and \ref{sec4} we exemplify the theory
discussed above by considering respectively the case of  stochastic
lattice gases, giving also a microscopic definition of work, and Langevin dynamics.

\section{Basic Assumptions}
\label{s:2}

We introduce in this section the thermodynamic description of out of
equilibrium driven diffusive systems which are characterized by
conservation laws. For simplicity of notation, we restrict to the case
of a single conservation law, e.g.\ the conservation of the mass.  The
system is in contact with boundary reservoirs, characterized by their
chemical potential $\lambda$, and under the action of an external
field $E$.  We denote by $\Lambda \subset \bb R^d$ the bounded region
occupied by the system, by $x$ the macroscopic space coordinates and
by $t$ the macroscopic time.  With respect to our previous work
\cite{primo,mindiss,curr,rev,tow,lagpt}, we consider the case in which
$\lambda$ and $E$ can depend explicitly on the time $t$.

The macroscopic dynamics is given by the hydrodynamic equation for the
density which satisfies the following general assumption, based on the
notion of local equilibrium.  It will be convenient to use a different
notation for space-time density paths and space dependent density
profiles. In the sequel we denote by $u=u(t,x)$ space-time
dependent paths and by $\rho=\rho(x)$ time independent profiles.

\begin{itemize}
\item[1.]  \emph{The macroscopic state is completely described by the
    local density $u(t,x)$ and the associated current $j(t,x)$.}

\item[2.]\emph{The macroscopic evolution is given by the continuity equation
together with the constitutive equation which express the current in
function of the density. Namely,
\begin{equation}
\label{2.1}
\begin{cases}
\partial_t u (t) + \nabla\cdot j (t) = 0,\\
j (t)= J(t,u(t)),
\end{cases}
\end{equation}
where we omit the explicit dependence on the space variable $x\in\Lambda$.
For driven diffusive systems the constitutive equation takes the form
\begin{equation}
\label{2.2}
J(t,\rho)  = - D(\rho) \nabla\rho + \chi(\rho) \, E(t)
\end{equation}
where the \emph{diffusion coefficient} $D(\rho)$ and the \emph{mobility}
$\chi(\rho)$ are $d\times d$ positive matrices.}

\item[3.]\emph{The transport coefficients $D$ and $\chi$ satisfy the
    local Einstein relation
\begin{equation}
\label{ein_rel}
D(\rho) = \chi(\rho) \, f''(\rho),
\end{equation}
where $f$ is the equilibrium free energy per unit of volume.}

\item[4.]\emph{The equations \eqref{2.1}--\eqref{2.2} have to be
    supplemented by the appropriate boundary condition on
    $\partial\Lambda$ due to the interaction with the external
    reservoirs. If $\lambda(t,x)$, $x\in\partial \Lambda$, is the
    chemical potential of the external reservoirs, this boundary
    condition reads
\begin{equation}
\label{2.3}
f'\big(u(t,x) \big) = \lambda(t,x), \qquad x\in\partial \Lambda.
\end{equation}
}
\end{itemize}

In the case of stochastic microscopic models with time independent
driving, the above macroscopic description is derived in the diffusive
scaling limit \cite{primo,curr,dd,KL,S}. As we discuss later, the
extension to time dependent driving is straightforward.

Given time-independent chemical potential $\lambda(x)$ and external
field $E(x)$, we drop the dependence on $t$ from $J(t,\rho)$ and
denote by $\bar\rho_{\lambda, E}$ the stationary
solution of \eqref{2.1}--\eqref{2.3}, 
\begin{equation}
\label{05}
\begin{cases}
\nabla \cdot J(\bar\rho)= \nabla \cdot \Big( -D(\bar\rho)
\nabla\bar\rho + \chi(\bar\rho) \, E  \Big) = 0,  & x\in\Lambda, \\ 
 f' (\bar\rho(x)) = \lambda (x), & x\in\partial \Lambda.   
\end{cases}
\end{equation}

Observe that if the field $E$ is gradient, $E=\nabla U$, and if it is
possible to choose the arbitrary constant in the definition of $U$
such that $U(x)=\lambda(x)$, $x\in\partial \Lambda$, then the
stationary solution satisfies
$f'\big(\bar\rho_{\lambda,E}(x)\big)=U(x)$ and the stationary current
vanishes, $J(\bar\rho_{\lambda,E})=0$.  Conversely, given any profile
$\bar\rho(x)$ it is possible to choose $\lambda(x)$ and $E(x)$ so that
$\bar\rho$ solves \eqref{05} and moreover $J(\bar\rho)=0$. It is
indeed enough to set $\lambda(x) = f'(\bar\rho(x))$, $x\in
\partial\Lambda$, and $E(x) = \nabla f'(\bar\rho(x))$, $x\in\Lambda$.
According to the point of view introduced in \cite{tow}, we refer to
this case as (inhomogeneous) equilibrium states.

Given time-dependent chemical potential $\lambda(t,x)$ and external
field $E(t,x)$, for $t\ge 0$ the profile $\bar\rho_{\lambda(t),E(t)}$
is the solution of \eqref{05} with $\lambda$ and $E$ ``frozen'' at the
time $t$. By using such profile, it is possible to reduce the
equations with time-dependent boundary conditions \eqref{2.3} to the
case of time independent boundary conditions. Indeed, by writing
$u(t)=\bar\rho_{\lambda(t),E(t)} +v(t)$ we deduce that $v$ solves
\begin{equation*}
\partial_t v = \nabla \cdot 
\big[ D\big(\bar\rho_{\lambda(t),E(t)} +v \big ) 
\nabla \big(\bar\rho_{\lambda(t),E(t)}+v\big) - 
\chi\big( \bar\rho_{\lambda(t),E(t)}+ v \big)E \big] - \partial_t
\bar\rho_{\lambda(t),E(t)} 
\end{equation*} 
with boundary conditions $v(t,x)=0$ for $x \in \partial \Lambda$.

\subsection*{Energy balance}

The energy exchanged between the system and the external
reservoirs and fields in the time interval $[0,T]$ is given by 
\begin{equation}
\label{W=}
\int_{0}^{T}\! dt \, \Big\{ 
- \int_{\partial\Lambda} \!d\sigma(x) \: \lambda (t,x) \: j(t,x) \cdot
\hat{n}(x)  +\int_\Lambda \!dx\: j(t,x) \cdot E(t,x) \Big\},
\end{equation}
where $\hat n$ is the outer normal to $\partial \Lambda$ and $d\sigma$
is the surface measure on $\partial \Lambda$.  The first term on the
right hand side is the energy provided by the reservoirs while the
second is the energy provided by the external field. 

Fix time dependent paths $\lambda(t,x)$ of the chemical potential and
$E(t,x)$ of the driving field. Given a density profile $\rho$, let
$u(t, x)$, $j(t,x)$, $t \ge 0$, $x\in\Lambda$, be the solution of
\eqref{2.1}--\eqref{2.3} with initial condition $\rho$.  We then
denote by $W_{[0,T]} = W_{[0,T]}({\lambda, E, \rho})$, the energy
exchanged between the system and the external driving, dropping the
subscript when $T=+\infty$. We claim that
\begin{equation}
\label{03}
W_{[0,T]}  \ge  F (u(T)) - F(\rho),
\end{equation} 
where $F$ is the equilibrium free energy functional,
\begin{equation}
\label{10}
F(\rho) = \int_\Lambda \!dx \: f (\rho(x)).
\end{equation}
Indeed, by using the boundary condition \eqref{2.3} and by the
divergence theorem in \eqref{W=}, (from now on we drop from the notation the
dependence on $x$)
\begin{equation}
\label{04}
\begin{split}
&W_{[0,T]}  = 
\int_{0}^{T}\! dt \, \Big\{ 
- \int_{\partial\Lambda} \!d\sigma \, f'(u(t)) \, j(t) \cdot \hat{n}  
+\int_\Lambda \!dx\, j(t) \cdot E(t) \Big\}
\\
&\quad= \int_{0}^{T} \!dt \int_\Lambda \!dx \, \big\{ 
- \nabla\cdot \big[ f'(u(t) ) \, j(t) \big] +  j (t) \cdot E (t)\big\}
\\ 
&\quad= \int_{0}^{T} \!dt \int_\Lambda \!dx \, \big[ 
   - f'(u(t)) \nabla \cdot j(t)  - f''(u(t)) \nabla u(t)
   \cdot j(t) + j(t)\cdot E(t) \big]  
\\
&\quad= \int_{0}^{T} \!dt  \, \frac{d}{dt}  \int_\Lambda \!dx  \,
f( u(t) ) \;+\; \int_{0}^{T} \!dt  \int_\Lambda \!dx \;
j(t)\cdot \chi(u(t) )^{-1} j(t),
\end{split}
\end{equation}
where we used the continuity equation \eqref{2.1}, the Einstein
relation \eqref{ein_rel}, and the constitutive equation \eqref{2.2}.
Since the first term is a total derivative and the second one is
positive, the inequality \eqref{03} follows.  

This argument provides a dynamic derivation of the second law of
thermodynamics as expressed by the Clausius inequality \eqref{03}. The
key ingredients have been the assumption of local equilibrium together
with the local Einstein relationship \eqref{ein_rel}.

\section{Equilibrium states}
\label{s:es}

We examine in this section the case of equilibrium states and their
transformations.

\subsection*{Reversible and quasi static transformations}

We consider first the simpler case of spatially homogeneous equilibrium
states.  Such states are characterized by a vanishing external field
$E$ and by a chemical potential $\lambda$ constant in space and time.
In this case the stationary solution $\bar \rho_{\lambda, 0}$ of the
hydrodynamic equations \eqref{2.1}--\eqref{2.3} is the constant $\rho$
satisfying $f'(\rho) = \lambda$. Hereafter, we denote $\bar
\rho_{\lambda, 0}$ simply by $\bar \rho_{\lambda}$.

Fix two constant chemical potentials $\lambda_0$, $\lambda_1$.
Consider a system initially in the state $\bar\rho_0 =
\bar\rho_{\lambda_0}$ which is driven to a new state $\bar\rho_1 =
\bar\rho_{\lambda_1}$ by changing the chemical potential in
time in a way that $\lambda(t)=\lambda_0$ for $t\le 0$ and
$\lambda(t)=\lambda_1$ for $t\ge T$; here $T$ is some fixed positive
time.   
This transformation from $\bar\rho_0$ to $\bar\rho_1$ is
called \emph{reversible} if the energy exchanged with the reservoirs is minimal. 
A basic thermodynamic principle asserts that reversible transformation
are accomplished by a sequence of equilibrium states and are well
approximated by \emph{quasi static} transformations, transformations
in which the variation of the chemical potential is very slow so that 
the density profile at time $u(t)$ is very close to the stationary
profile $\bar\rho_{\lambda(t)}$.
We show that this principle can be derived from the general
assumptions of Section~\ref{s:2}.

Let $u(t, x)$, $j(t,x)$, $t \ge 0$, $x\in\Lambda$, be the solution of
\eqref{2.1}--\eqref{2.3} with initial condition $\bar\rho_0$. 
Since the chemical potential is equal to $\lambda_1$ for $t\ge T$, 
it holds $u(t)\to\bar\rho_1$ as $t\to+\infty$. Moreover, as
$\bar\rho_1$ is an equilibrium state, the current $j(t)$ relaxes 
to $J(\bar\rho_1)=0$. Observe that, since the system has a finite
relaxation time, the convergence is  exponentially fast.  
We deduce that the last integral in \eqref{04} is finite as
$T\to\infty$ and 
\begin{equation}
\label{11}
\begin{split}
W  & =   \int_{0}^{\infty} \!dt  \, \frac{d}{dt}  \int_\Lambda \!dx  \,
f( u(t) ) \;+\; \int_{0}^{\infty} \!dt  \int_\Lambda \!dx \;
j(t)\cdot \chi(u(t) )^{-1} j(t) 
\\
& \ge |\Lambda| \, \big[ f (\bar\rho_1) -f(\bar\rho_0) \big]. 
\end{split}
\end{equation}
Note that we did not assume any regularity
of the chemical potential in time so that it can be also
discontinuous. 

It remains to show that in the quasi static limit 
equality in \eqref{11} is achieved. 
That is the thermodynamic relation
\begin{equation}
\label{12}
W =  \Delta F
\end{equation}
holds, where $\Delta F = |\Lambda| \big[ f(\bar\rho_1) - f(\bar\rho_0)
\big]$ is the variation of the free energy. 
If this is case, by running the transformation backward in time, we can return to the
original state exchanging the energy $-\Delta F$.   
For this reason the transformations for which \eqref{11} becomes
equality are called reversible.
Since for any fixed transformation the inequality in
\eqref{11} is strict because the second term on the right hand side of
the first line in \eqref{11} cannot be identically zero, reversible
transformations cannot be achieved exactly. 
We can however exhibit a sequence of transformations for which the second term
on the right hand side of the first line in \eqref{11} term can be
made arbitrarily small.  This sequence of transformations is what we
call quasi static transformations.  
Fix a smooth function $\lambda(t)$ such that $\lambda(0)=\lambda_0$
and $\lambda(t)=\lambda_1$ for $t\ge T$. Given $\delta >0$ we set
$\lambda_\delta (t) = \lambda (\delta t)$.  Since $E=0$, the second
term on the right hand side of \eqref{11} is given by
\begin{equation*}
\int_{0}^{\infty}\!dt \int_\Lambda\!dx \, 
\nabla f' (u_\delta(t))  \cdot 
\chi(u_\delta(t)) \nabla f' (u_\delta(t)),
\end{equation*}
where $u_\delta$ is the solution to \eqref{2.1}--\eqref{2.3} with
initial condition $\bar\rho_0$ and boundary conditions
$\lambda_\delta(t)$.  Recall that $\bar\rho_{\lambda_\delta(t)}$
is the equilibrium state associated 
to the constant chemical potential $\lambda_\delta (t)$ (with $t$
frozen).  
Since $\nabla  f' (\bar\rho_{\lambda_\delta(t)}) =0$, we can
rewrite the previous integral as
\begin{equation*}
\int_{0}^{\infty}\!dt \int_\Lambda \!dx\, 
\nabla \big[ f' (u_\delta(t)) - f' (\bar\rho_{\lambda_\delta(t)}) \big] 
\cdot \chi(u_\delta(t)) 
\nabla \big[ f' (u_\delta(t)) - f' (\bar\rho_{\lambda_\delta(t)} ) \big].
\end{equation*}
The difference between the solution of the hydrodynamic equation
$u_\delta(t)$ and the stationary profile $\bar\rho_{\lambda_\delta(t)}$ is
of order $\delta$ uniformly in time, and so is the difference $f'
(u_\delta(t)) - f' (\bar \rho_{\lambda_\delta(t)})$.  As the integration
over time essentially extends over an interval of length
$\delta^{-1}$, the previous expression vanishes for $\delta\to 0$.
This implies that equality in \eqref{11} is achieved in the limit
$\delta\to 0$. 
Note that in the
previous argument we did not use any special property of the path
$\lambda(t)$ besides its smoothness in time. The trajectory $\lambda
(t)$ from $\lambda_0$ to $\lambda_1$ can be otherwise arbitrary.

\medskip
We now discuss the case of spatially inhomogeneous equilibrium states.
According to the point of view introduced in \cite{tow}, in absence of
external magnetic fields, such states $\bar\rho=\bar\rho(x)$ are
characterized by the vanishing of the associated current,
$J(\bar\rho)=0$.  An example is provided by a sedimentation
equilibrium in gravitational and centrifugal fields.

Consider a density profile $\rho$, a time dependent chemical potential
$\lambda (t,x)$ and a time dependent external field $E(t,x)$. We
assume that $\lambda (t,x)$, $E(t,x)$ converge to $\lambda_1 (x)$,
$E_1(x)$ as $t\to+\infty$ fast enough, e.g.\ exponentially fast. 
Let $\bar\rho_1 = \bar \rho_{\lambda_1, E_1}$ be the stationary state
associated to the chemical potential $\lambda_1$ and the external
field $E_1$. We also assume that $\bar\rho_1$ is an equilibrium state,
that is the current $J(\bar\rho_1)$ vanishes.

Let $u(t, x)$, $j(t,x)$, $t \ge 0$, $x\in\Lambda$, be the solution of
\eqref{2.1}--\eqref{2.3} with initial condition $\rho$. Since
$\bar\rho_1$ is an equilibrium state, the current $j(t)$ relaxes as
$t\to +\infty$, to $J(\bar\rho_1)=0$.  The argument presented for
homogeneous equilibrium applies also to the present setting and yields
\begin{equation}
\label{06b}
W(\lambda, E, \rho)  \ge  F (\bar\rho_1) - F(\rho),
\end{equation}
where $F$ is the equilibrium free energy defined in \eqref{10}. 

It remains to introduce quasi static transformations in this more
general context and show that equality in \eqref{06b} is achieved. 
Let $\lambda_0 (x) = \lambda (0,x)$, $E_0(x) = E(0,x)$. Assume that
the initial profile $\rho$ is the stationary profile associated to
$\lambda_0$, $E_0$, $\rho = \bar\rho_{\lambda_0, E_0}=\bar\rho_0$, and that
$\bar\rho_0$, is an equilibrium state, $J(\bar \rho_{0}) =0$.
Fix $T>0$ and choose smooth functions
$(\lambda(t), E(t))$, such that $(\lambda(0), E(0)) =
(\lambda_0, E_0)$, $(\lambda(t), E(t)) = (\lambda_1, E_1)$, $t\ge T$, 
and $J(\bar\rho_{\lambda(t), E(t)})=0$ for $t\ge 0$.
Such transformations always exist but are not unique. We may, for
instance, first choose a smooth path $\bar\rho(t)$, 
such that $\bar\rho (0)=\bar\rho_0$ and $\bar\rho(t)=\bar\rho_1$ for
$t\ge T$.
Then choose $\lambda(t)=f'(\bar\rho(t))$ and $E(t)= \nabla
f'(\bar\rho(t))$.  In view of the discussion below \eqref{05}, we then
have $\bar\rho_{\lambda(t),E(t)}= \bar\rho(t)$.  For $\delta >0$ ,
let $(\lambda_\delta (t), E_\delta (t)) = (\lambda (\delta t),
E(\delta t))$. Let $u_\delta(t)$ be the solution of
\eqref{2.1}--\eqref{2.3} with initial condition $\bar\rho_0$, boundary
condition $\lambda_\delta (t)$ and external field $E_\delta (t)$.
At this point we can repeat the argument for homogeneous equilibrium
states and show that equality in \eqref{06b} is achieved in the
quasi static limit $\delta\to 0$.

\subsection*{Excess work}

Consider a transformation $(\lambda (t), E(t))$, $t\ge 0$, and an
initial density profile $\rho$.  We assume that as $t\to +\infty$ it
holds $(\lambda (t), E(t))\to (\lambda_1,E_1)$ fast enough where
$(\lambda_1,E_1)$ defines the equilibrium state
$\bar\rho_1=\bar\rho_{\lambda_1,E_1}$, i.e.\ $J(\bar\rho_1)=0$.  We
then introduce the \emph{excess work} $W_\mathrm{ex} =
W_\mathrm{ex}(\lambda,E,\rho)$ as the difference between the energy
exchanged between the system and the external driving and the work
involved in a reversible transformation from $\rho$ to $\bar\rho_1$,
namely
\begin{equation}
\label{exwork} 
W_\mathrm{ex} = W(\lambda, E, \rho)  - \min W = 
  \int_0^\infty \!dt \int_{\Lambda} \!dx \, j(t) \cdot
  \chi(u(t))^{-1} j(t) , 
\end{equation}
where we used \eqref{11} as well as the fact that the minimum of $W$ is given by
the right hand side of \eqref{12}. 
Observe that $W_\mathrm{ex}$ is a positive functional of the transformation
$(\lambda(t), E(t))$ and the initial condition $\rho$. Of course, by taking a 
sequence of quasi static transformations $W_\mathrm{ex}$ can be made
arbitrarily small. 
Below we shall compute $W_\mathrm{ex}$ for specific transformations
and illustrate its thermodynamic relevance.

\subsection*{Relaxation path  and availability}

Consider an equilibrium system in the state $\bar\rho_0$,
characterized by a chemical potential $\lambda_0$ and an external
field $E_0$. This system is put in contact with reservoirs at constant
chemical potential $\lambda_1$ and an external field $E_1$, different
from the chemical potential $\lambda_0$ and the external field $E_0$
associated to $\bar\rho_0$. For $t> 0$ the system thus evolves
according to the hydrodynamic equation \eqref{2.1}--\eqref{2.3} with
initial condition $\bar\rho_0$, external field $E_1$, and boundary
condition $\lambda_1$. Such a transformation can be realized by
considering first a smooth transition from $\lambda_0$ to
$\lambda_1$ and then taking the limit in which it becomes a step function. 
When $t\to +\infty$ the system relaxes to the equilibrium state
$\bar\rho_1$.  
In view of \eqref{exwork} and the constitutive equation \eqref{2.2},
the excess work along such a path is given by
\begin{equation*}
W_\mathrm{ex} (\lambda_1, E_1, \bar\rho_0) 
= - \int_0^{\infty}\!dt \int_\Lambda \!dx \: 
\big[ \nabla f'(u(t)) - E_1 \big] \cdot J(u(t)) . 
\end{equation*}
Since $J(\bar\rho_1)=0$, $\nabla f'(\bar\rho_1) = E_1$, and we may
replace $E_1$ by $\nabla f'(\bar\rho_1)$ in the previous equation. As
$u(t)$ and $\bar\rho_1$ satisfy the same boundary conditions, after an
integration by parts the previous expression becomes
\begin{equation*}
\begin{split}
W_\mathrm{ex} (\lambda_1, E_1, \bar\rho_0) 
& = \int_0^{\infty}\!dt \int_\Lambda \!dx \: 
\big[ f'(u(t)) - f'(\bar\rho_1) \big] \nabla \cdot J(u(t))   \\
& = - \int_0^{\infty}\!dt \int_\Lambda \!dx \: 
\big[ f'(u(t)) - f'(\bar\rho_1) \big] \, \partial_t u(t).    
\end{split}
\end{equation*}
We have therefore shown that
\begin{equation}
\label{exw=qp}
W_\mathrm{ex} (\lambda_1, E_1, \bar\rho_0) = \int_\Lambda \!dx\: 
\big[ f(\bar\rho_0) - f(\bar\rho_1) -  
f'(\bar\rho_1) \big( \bar\rho_0 - \bar\rho_1\big) \big].
\end{equation}
Observe that the excess work $W_\mathrm{ex}$ is not the difference of
a thermodynamic potential between the states $\bar\rho_0$ and
$\bar\rho_1$.  
In the case of spatially homogeneous equilibria with vanishing external
field, \eqref{exw=qp} becomes
\begin{equation}
\label{13}
W_\mathrm{ex}[\lambda_1, 0, \bar\rho_0] = |\Lambda|
\big[ f(\bar\rho_0) - f(\bar\rho_1) -  
\lambda_1 \big( \bar\rho_0 - \bar\rho_1\big)\big]. 
\end{equation}

To connect this computation with classical thermodynamics, we briefly 
recall the notion of \emph{availability}, see e.g., \cite[Ch.~7]{pippard}. 
Since the temperature of the system is the same of the environment, the
availability per unit of volume is defined  by 
$a= f(\bar\rho_0) -\lambda_1 \bar\rho_0$.
The function $a$, which depends on the state of the system
$\bar\rho_0$ and the environment $\lambda_1$, can be used to compute
the maximal useful work that can be extracted from the system in the
given environment. More precisely, recalling that
$f'(\bar\rho_1)=\lambda_1$,
\begin{equation}
\label{da}
-\Delta a = f(\bar\rho_0) - f(\bar\rho_1) - \lambda_1 (\bar\rho_0
-\bar\rho_1)\ge 0
\end{equation}
is the the maximal useful work per unit of volume
that can be extracted from the system in the given environment, see 
\cite[Ch.~7]{pippard} or \cite[\S~20]{LL}.  The inequality in
\eqref{da} is due to the convexity of $f$ and expresses the
thermodynamic stability.
We have thus concluded that, along the relaxation path specified above
the excess work $W_\mathrm{ex}$ is equal to the maximal useful work
that can be extracted from the system.

\subsection*{Fluctuations and quasi potential}

The Einstein theory of thermodynamic fluctuations, see e.g.\
\cite[Eq.~(112.2)]{LL}, establishes a precise connection between the
excess work computed along the transformation described before and the
probability of observing a fluctuation.  Denote by $\mu^{\lambda, E}$
the statistical ensemble of an equilibrium (not necessarily spatially
homogeneous) system in contact with reservoirs at chemical potential
$\lambda$ and with an external field $E$.  The probability of
observing a fluctuation $\rho$ of the density in the macroscopic
volume $\Lambda$ can be expressed as
\begin{equation}
\label{ld}
\mu^{\lambda, E} (\rho_\epsilon \approx \rho) 
\asymp \exp\big\{ - \epsilon^{-d} \, \beta \, V_{\lambda, E}(\rho)\big\}, 
\end{equation}
where $\beta=1/\kappa T$ (here $T$ is the temperature), 
$\epsilon\ll 1$ is the (a-dimensional) scaling factor, i.e.\ the ratio
between the microscopic length scale (say the typical intermolecular
distance) and the macroscopic one, and  $\rho_\epsilon$ is the \emph{empirical
  density} namely, $\rho_\epsilon(x)$ is the average number of
particles is a macroscopically small volume around $x$.  The symbol
$\asymp$ denotes logarithmic equivalence as $\epsilon \to 0$ and
\begin{equation}
\label{14}
V_{\lambda, E}(\rho) = W_\mathrm{ex}(\lambda,E,\rho). 
\end{equation}
In the right hand side of \eqref{14} the chemical potential $\lambda$
and the external field $E$ are constant in time so that
$W_\mathrm{ex}$ is given by \eqref{exw=qp}.

Referring to \cite{mindiss,tow} for more details, we briefly present
the connection of the functional $V_{\lambda, E}$ to a control
problem.
Instead of computing the asymptotic probability of observing a given
fluctuation, we take an active viewpoint looking at the most
convenient way to produce such fluctuation.  Consider at time
$t=-\infty$ an equilibrium system in the state $\bar\rho_1$ in contact
with reservoirs whose chemical potential is $\lambda_1$ and an
external field $E_1$.  We drive the system in the time interval
$(-\infty, 0]$ to the new state $\rho$, attained at time $t=0$ by
superimposing a field $e(t)$ to the original external field $E_1$.  We
introduce the associated cost functional $I$ as
\begin{equation}
  \label{Idyn}
  I(u,j) = \frac 14 \int_{-\infty}^0\!dt \int_{\Lambda}\! dx \: 
            e(t) \cdot \chi(u(t)) e(t)   
\end{equation}
where the path $(u(t),j(t))$, $t\in (-\infty,0]$ satisfies
\eqref{2.1}--\eqref{2.3} with fixed chemical potential $\lambda_1$ and
external field $E_1+e(t)$. 
Observe that there is a one-to-one correspondence between the path
$(u(t),j(t))$ and the driving field $e(t)$. We can thus consider, as in
\eqref{Idyn}, the functional $I$ to be defined on the set of space-time paths
$(u(t),j(t))$. 

As discussed below \eqref{05}, the arbitrary density profile $\rho$
can be regarded as an equilibrium state associated to some chemical potential
and some external field. 
As shown in  \cite{mindiss,tow}
\begin{equation}
\label{vvp}
V_{\lambda_1, E_1} (\rho) = W_\mathrm{ex}(\lambda_1,E_1,\rho) = 
\min  I (u,j),
\end{equation}
where the minimum is carried over all driving $e(t)$ such that
$u(0)=\rho$. Observe that in this argument $(\lambda_1,E_1)$ is the
state at time $t=-\infty$ while $\rho$ is the density profile at time $t=0$.

As shown in \cite{tow}, the optimal trajectory $(u(t),j(t))$ for the
variational problem on the right hand side of \eqref{vvp} is the time
reversal of the relaxation trajectory defined as follows. It is the
solution to the hydrodynamic equations \eqref{2.1}--\eqref{2.3} where
the chemical potential and the external field are respectively equal
to $\lambda_1$ and $E_1$, while the initial condition, at $t=0$, is
$\rho$. In particular it relaxes toward $\bar\rho_1$. A simple
computation indeed shows that if we evaluate the functional $I$ along
the time reversal of such trajectory we indeed get the excess work
$W_\mathrm{ex}(\lambda_1,E_1,\rho)$ that has been computed in before.
Such a time reversal symmetry is a peculiar feature of equilibrium
states.

As discussed in \cite{primo,curr}, the functional $I(u,j)$ describes
the probability of space-time fluctuations of the density and current
and, by solving the variational problem on the right hand side of
\eqref{vvp}, the probability of static fluctuations \eqref{ld} is
recovered.  In the concrete models of stochastic lattice gases, these
statements can be rigorously proven.

\subsection*{Relative entropy}

We conclude this section establishing the connection between the
functional $V_{\lambda, E}$ with the Gibbs states of equilibrium
statistical mechanics. For simplicity of notation we consider the case
of lattice gases without external field and constant chemical
potential, i.e.\ the case of homogeneous equilibrium states. 
Let $\Lambda_\ell$ be the cube of side length $\ell$ in
$\bb Z^d$ and, for $\lambda\in \bb R$, let $\mu_{\ell}^\lambda$ be
the grand-canonical Gibbs measure on $\Lambda_\ell$ with chemical
potential $\lambda$,
\begin{equation}
\label{gibbs} 
\mu_\ell^\lambda (\eta) = \frac {1}{Z_\ell(\lambda)} 
\exp\Big\{ - \beta H_\ell (\eta) +
\beta \lambda \sum_{x\in\Lambda_\ell} \eta(x) \Big\},
\end{equation}
where $\beta=1/\kappa T$, $\eta(x)$, $x\in\Lambda_\ell$, are the
occupation variables, $H_\ell (\eta)$ is the energy of the
configuration $\eta$, and $Z_\ell (\lambda)$ is the grand-canonical
partition function. 
The pressure $p$ is given by 
\begin{equation}
  \label{press}
p(\lambda) =\ \frac 1\beta \, \lim_{\ell \to \infty} \frac 1{\ell^{d}} 
\log Z_\ell(\lambda),
\end{equation}
and the free energy per unit of volume $f$, the function which appears in
\eqref{ein_rel}, is obtained as the Legendre transform of $p$,
\begin{equation*}
f(\rho) =\sup_\lambda \big\{ \rho \lambda - p(\lambda) \big\}.
\end{equation*}

The \emph{relative entropy} $S(\nu|\mu)$ of the probability $\nu$ with
respect to $\mu$ is defined by
\begin{equation}
\label{relen}
S(\nu|\mu) = \int \! d\mu \: \frac{d\nu}{d\mu} \log
\frac{d\nu}{d\mu}\; \cdot
\end{equation}
Fix two chemical potentials $\lambda_0$ and $\lambda_1$. We claim that 
\begin{equation}
\label{ler}
\lim_{\ell\to \infty} \, \frac 1{\ell^{d}} 
S\big( \mu_{\ell}^{\lambda_0} \big|\mu_{\ell}^{\lambda_1}\big) 
= \beta \big[ 
f(\bar\rho_0) - f(\bar\rho_1) - \lambda_1 (\bar\rho_0-\bar\rho_1)
\big], 
\end{equation}
where $\bar\rho_0$ and $\bar\rho_1$ are the densities associated to
$\lambda_0$ and $\lambda_1$.  In view of \eqref{13}--\eqref{14} this
implies that in the thermodynamic limit $\ell\to \infty$ the relative
entropy per unit of volume is proportional to the function $
V_{\lambda_1,0}(\bar\rho_0)$ per unit volume.  To prove \eqref{ler},
observe that in view of \eqref{relen} and the Gibbsian form
\eqref{gibbs},
\begin{equation*}
\frac 1 {\ell^d} \, 
S \big( \mu_{\ell}^{\lambda_0} \big| \mu_{\ell}^{\lambda_1} \big)
= \frac 1{\ell^d}  \log \frac{ Z_{\ell}(\lambda_1)} {Z_\ell(\lambda_0)}
+ \beta (\lambda_0 -\lambda_1) 
\sum_{\eta} \mu_{\ell}^{\lambda_0} (\eta) 
\: \frac 1{\ell^d} \sum_{x\in \Lambda_\ell} \eta(x) 
\end{equation*}
By definition of the pressure, the first term converges to
$\beta [ p(\lambda_1) - p(\lambda_0)]$, while the second one converges to
$\beta (\lambda_0-\lambda_1)\bar\rho_0$.  The identity \eqref{ler} then
follows by Legendre duality.

The above interpretations of the functional $V_{\lambda,E}$, hereafter 
referred to as the \emph{quasi potential}, reveal
the connections between the static and dynamical properties of
equilibrium systems. These connections are the starting point
for a macroscopic description of nonequilibrium systems.

\section{Nonequilibrium states}   
\label{s:4}

Nonequilibrium states are characterized by the presence of a non
vanishing current in the stationary density profile. Therefore, to
maintain such states one needs to dissipate a positive amount of energy
per unit of time.  If we consider a transformation between
nonequilibrium stationary states, the energy dissipated along such
transformation will necessarily include the contribution needed to
maintain such states. The arguments of the previous section have
therefore to be modified in order to take into account this amount of
energy.  This issue, first raised in \cite{op}, has been more recently
considered e.g.\ in \cite{bmw,maes2,hs,kn,knst}.

The appropriate definition of thermodynamic functionals for
nonequilibrium systems is a central but difficult topic.  Our starting
point is the fluctuation formula \eqref{ld}, which, provided we
replace $\mu^{\lambda,E}$ with the appropriate ensemble, makes good
sense also in nonequilibrium so that the notion of the quasi potential
can be defined also for nonequilibrium states.  This has been the
basis of our previous work on the subject \cite{primo,rev}. We recall
however that even for equilibrium systems the quasi potential is not
really a function of the state but expresses a property of the system
in a given environment, see \eqref{exw=qp}.  
In this section we show that - even for nonequilibrium states - 
the quasi potential is connected to the excess work 
and to the specific relative entropy between two states.  
We first recall some relevant results from \cite{primo,mindiss}.

\subsection*{Quasi potential}

Fix time independent chemical potential $\lambda=\lambda(x)$,
$x\in\partial\Lambda$, external field $E=E(x)$, $x\in\Lambda$, and
recall that $\bar\rho_{\lambda,E}$, the solution of \eqref{05}, is the
stationary solution of the hydrodynamic equation.  We assume that
$\lambda,E$ define a nonequilibrium state in the sense that
$J(\bar\rho_{\lambda,E})\neq 0$. The statistical ensemble associated
to such state is still denoted by $\mu^{\lambda,E}$. Then, as shown in
\cite{primo}, the fluctuation formula \eqref{ld} holds where the quasi
potential $V_{\lambda,E}$ solves the same variational problem as in
equilibrium states. Namely,
\begin{equation}
\label{19}
  V_{\lambda,E} (\rho) = \min I(u,j)
\end{equation}
where $I$ is the action functional defined in \eqref{Idyn} and the
minimum is carried out over all paths such that
$u(-\infty)=\bar\rho_{\lambda,E}$ and $u(0)=\rho$.

In nonequilibrium there is no simple formula for the quasi potential
but it can be characterized \cite{primo,mindiss,curr} as the maximal solution
of the stationary Hamilton-Jacobi equation
\begin{equation}
\label{30}
\int_{\Lambda} \!dx\, \nabla  
\frac {\delta V_{\lambda,E}(\rho)}{\delta\rho} \cdot \chi(\rho) \, \nabla  
\frac {\delta V_{\lambda,E}(\rho)}{\delta\rho} \;-\; \int_{\Lambda}
\!dx\, \frac {\delta V_{\lambda,E}(\rho)}{\delta\rho} \, \nabla \cdot J(\rho)
\;=\; 0.
\end{equation}
where ${\delta V_{\lambda, E}}/{\delta \rho}$ vanishes at the boundary
$\partial \Lambda$ and $\rho$ satisfies the boundary condition
$f'(\rho(x)) = \lambda(x)$, $x\in\partial\Lambda$.  The current
$J(\rho)$ in \eqref{2.2} may therefore be decomposed as
\begin{equation}
\label{decur}
J(\rho) = J_{\mathrm{S}}(\rho) +
J_{\mathrm{A}}(\rho),
\end{equation}
where 
\begin{equation}
\label{jd}
J_{\mathrm{S}}(\rho)= 
-\chi(\rho) \, \nabla  
\frac {\delta V_{\lambda,E}(\rho)}{\delta\rho}
\end{equation}
and $J_{\mathrm{A}}(\rho) = J(\rho) - J_{\mathrm{S}}(\rho)$.  In view
of the stationary Hamilton-Jacobi equation \eqref{30}, the
decomposition \eqref{decur} is orthogonal in the sense that for each
$\rho$
\begin{equation}
\label{ort}
\int_{\Lambda}\!dx \: J_{\mathrm{S}}(\rho) \cdot \chi(\rho)^{-1}
J_{\mathrm{A}} (\rho) = 0 \;.
\end{equation}
We shall refer to $J_{\mathrm{S}}(\rho)$ as the \emph{symmetric}
current and to $J_{\mathrm{A}}(\rho)$ as the \emph{antisymmetric}
current. This terminology refers to symmetric and antisymmetric part
of the underlying Markovian microscopic dynamics \cite{primo,mindiss,curr}. 
More precisely, the generator of the evolution can be decomposed into
a symmetric and an antisymmetric part which are respectively even and
odd under time reversal. 
The current $J_{\mathrm{S}}(\rho)$ is due to symmetric part and is
responsible for the relaxation, while $J_{\mathrm{A}}(\rho)$ to the 
antisymmetric part; we refer to Section~\ref{s:6.1} in which we discuss
this decomposition for the zero range model.
We finally emphasize that the decomposition \eqref{decur} depends not
trivially on $\lambda,E$.

Since the quasi potential $V_{\lambda,E}$ is minimal in the stationary
profile, we deduce that $J_{\mathrm{S}} (\bar\rho_{\lambda,E})=0$; 
namely, the stationary current is purely antisymmetric. 
In particular, $J_{\mathrm{A}} (\bar\rho_{\lambda,E})$ is the typical
current in the stationary nonequilibrium ensemble associated to
$(\lambda, E)$ and it is therefore experimentally accessible.
In view of the
general formula \eqref{04} for the total work, the amount of energy
per unit of time needed to maintain the system in the stationary
profile $\bar\rho_{\lambda,E}$ is
\begin{equation}
\label{toman}
\int_\Lambda \!dx \: 
J_\mathrm{A}(\bar\rho_{\lambda,E}) \cdot \chi(\bar\rho_{\lambda,E})^{-1} 
J_\mathrm{A}(\bar\rho_{\lambda,E}).
\end{equation}

\subsection*{Renormalized work} 

In view of the previous paragraph, by interpreting the ideas in
\cite{op}, it is natural to define in a nonequilibrium setting the
renormalized work as the total work minus the work needed to maintain the
stationary profile.  Fix, therefore, $T>0$, a density profile $\rho$,
and space-time dependent chemical potentials $\lambda(t)=\lambda(t,x)$
and external field $E(t)=E(t,x)$, $0\le t\le T$, $x\in\Lambda$.  Let
$u(t)=u(t, x)$, $j(t)=j(t,x)$, $t \ge 0$, $x\in\Lambda$, be the
solution of \eqref{2.1}--\eqref{2.3} with initial condition $\rho$.
Recalling \eqref{toman}, we thus define the
renormalized work $W^\textrm{ren}_{[0,T]} = 
W^\textrm{ren}_{[0,T]}(\lambda,E,\rho)$ performed by the
reservoirs and the external field in the time interval $[0,T]$ as
\begin{equation}
\label{Weff}
W^\textrm{ren}_{[0,T]} = W_{[0,T]} 
- \int_{0}^{T}\! dt \int_\Lambda \!dx\, J_\mathrm{A}(t,u(t)) \,
\cdot \chi(u(t))^{-1} J_\mathrm{A}(t,u(t)).
\end{equation}
In this formula $W_{[0,T]} =W_{[0,T]} (\lambda,E,\rho)$ is given in \eqref{W=},
\begin{equation*}
  J(t,\rho) =  J_\mathrm{S}(t,\rho) +J_\mathrm{A}(t,\rho),
  \qquad J_\mathrm{S}(t,\rho) = - \chi(\rho) \, \nabla  
  \frac {\delta V_{\lambda(t),E(t)}(\rho)}{\delta\rho}
\end{equation*}
in which $J(t,\rho)$ is given by \eqref{2.2} and 
$V_{\lambda(t),E(t)}$ is the quasi potential relative to
the state $\lambda(t), E(t)$ with frozen $t$.
Observe that the definition of the renormalized work involves the
antisymmetric current $J_\mathrm{A}(t)$ computed not at density
profile $\bar\rho_{\lambda(t), E(t)}$ but at the solution
$u(t)$ of the time dependent hydrodynamic equation. Therefore the
second term on the right hand side of \eqref{Weff} is not directly
measurable but requires first the computation of the quasi
potential. The thermodynamic relevance of the above definition will be
clear in the sequel.
In contrast with the terminology in \cite{op}, we used the term
\emph{renormalized work} for the functional in \eqref{Weff} in order to
reserve the term \emph{excess work} to the extra work with respect to
the minimal one.

Since the symmetric and the antisymmetric part of the current are
orthogonal, repeating the computation performed in \eqref{04}, we get
that
\begin{equation*}
W^\textrm{ren}_{[0,T]} (\lambda,E,\rho) = F(u(T)) - F(\rho) +
\int_{0}^{T}\! dt \int_\Lambda \!dx\, J_\mathrm{S}(t,u(t)) \cdot
\chi(u(t))^{-1} J_\mathrm{S}(t,u(t)).
\end{equation*}
We observe that for nonequilibrium states the quasi potential
is generically a non local functional in view of the long range
correlations \cite{rev,dd}. Therefore the symmetric current
$J_\mathrm{S}$ above is generically non local and thus the
renormalized work is also non local, compare with \cite{knst}.

Consider a density profile $\rho$ and a space-time dependent chemical
potential $\lambda (t)$ and external field $E(t)$.  Assume hereafter
that $\lambda (t), E(t)$ converge to $\lambda_1, E_1$ as $t\to+\infty$
fast enough, e.g.\ exponentially fast.  Let $\bar\rho_1 = \bar
\rho_{\lambda_1,E_1}$ be the stationary profile associated to the
chemical potential $\lambda_1, E_1$, and $(u(t),j(t))$, $t \ge 0$, be
the solution of \eqref{2.1}--\eqref{2.3} with initial condition
$\rho$.  Since $u(T)$ converges to $\bar\rho_1$, the symmetric part of
the current, $J_\mathrm{S}(u(T))$, relaxes as $T\to +\infty$ to
$J_\mathrm{S}(\bar\rho_1) = 0$ fast enough.
In particular, the last integral in the previous formula is convergent
as $T\to+\infty$ and we get 
\begin{equation}
\label{16}
W^\textrm{ren} (\lambda,E,\rho) =  F(\bar\rho_1) - F(\rho)
\;+\; \int_{0}^{\infty} \!dt  \int_\Lambda \!dx \:
J_\mathrm{S}(t,u(t)) \cdot \chi(u(t))^{-1} J_\mathrm{S}(t,u(t)) 
\end{equation}
where $F$ is the equilibrium free energy functional, see \eqref{10}.
In particular,
\begin{equation}
\label{15}
W^\textrm{ren}(\lambda, E,\rho)  \ge  F (\bar\rho_1) - F(\rho).
\end{equation}
As follows immediately from \eqref{Weff},
$W^\textrm{ren}_{[0,T]}(\lambda,E,\rho)\le W_{[0,T]} (\lambda,
E,\rho)$ and therefore \eqref{15} is stronger than the general
inequality \eqref{03}.
Indeed, \eqref{15} states that the Clausius inequality holds for the
renormalized work, see \cite{knst}.

\subsection*{Quasi static transformations}

As for equilibrium states, we show that, 
given two nonequilibrium states, there exists a sequence of 
transformations from the first to the second for which the second term
on the right hand of \eqref{16} can be made arbitrarily small.

Let $(\lambda_0, E_0) = (\lambda (0), E(0))$ and assume that the
initial profile $\rho$ is the stationary profile associated to
$\lambda_0, E_0$, i.e.\ $\rho = \bar\rho_{\lambda_0,E_0}$.  
Fix $T>0$ and choose smooth function $\lambda(t),E(t)$, $0\le
t\le T$, such that $(\lambda(0),E(0))=(\lambda_0,E_0)$, 
$(\lambda(T),E(T))=(\lambda_1,E_1)$.
For $\delta >0$, let $(\lambda_\delta (t),E_\delta(t)) = (\lambda
(\delta t),E(\delta t))$, and $(u_\delta(t),j_\delta(t))$ be the
solution of \eqref{2.1}--\eqref{2.3} with initial condition
$\bar\rho_0=\bar\rho_{\lambda_0,E_0}$, external field $E_\delta(t)$,
and boundary condition $\lambda_\delta (t)$.  The second term on the
right hand side of \eqref{16} is given by
\begin{equation*}
\int_{0}^{\infty}\!dt \int_\Lambda\!dx \, 
J_\mathrm{S} (t,u_\delta(t))  \cdot 
\chi(u_\delta(t))^{-1} J_\mathrm{S} (t,u_\delta(t)).
\end{equation*}
For each fixed $t$, let $\bar\rho_\delta(t) =
\bar\rho_{\lambda_\delta(t),E_\delta(t)}$ be the stationary profile 
associated to the driving $\lambda_\delta (t), E_\delta(t)$ with
frozen $t$.  
Since $J_\mathrm{S}(t,\bar\rho_\delta(t)) =0$, 
we can rewrite the previous integral as
\begin{equation*}
\int_{0}^{\infty}\!dt \int_\Lambda \!dx\, 
\big[ J_\mathrm{S} (t,u_\delta(t)) - 
J_\mathrm{S} (t,\bar\rho_\delta(t))\big] 
\cdot \chi(u_\delta(t))^{-1} 
\big[ J_\mathrm{S} (t,u_\delta(t)) - 
J_\mathrm{S} (t,\bar\rho_\delta(t))\big].
\end{equation*}
The difference between the solution of the hydrodynamic equation
$u_\delta(t)$ and the stationary profile $\bar\rho_\delta(t)$ is
of order $\delta$ uniformly in time, and so is the difference
$J_\mathrm{S} (t,u_\delta(t)) - J_\mathrm{S} (t,\bar\rho_\delta(t))$.  
As the integration over time essentially extends
over an interval of length $\delta^{-1}$, the previous expression
vanishes for $\delta\to 0$.  This implies that equality in
\eqref{15} is achieved in the limit $\delta\to 0$.
In this argument we did not use any special property of
the path $(\lambda(t),E(t))$ besides its smoothness in time, the
trajectory $(\lambda (t),E(t))$ from $(\lambda_0,E_0)$ to
$(\lambda_1,E_1)$ can be otherwise arbitrary.

Quasi static transformations thus minimize asymptotically the
renormalized work and in the limit $\delta\to 0$ we obtain the
nonequilibrium version of the thermodynamic relation \eqref{12}, that is
\begin{equation}
\label{17}
W^\textrm{ren}  =   \Delta F,
\end{equation}
where $\Delta F$ represents the variation of the equilibrium free
energy functional, $\Delta F = F(\bar\rho_1) - F(\bar\rho_0)$. 
It is remarkable that the Clausius inequality and the optimality of
quasi static transformations, basic laws of equilibrium
thermodynamics, admit exactly the same formulation, after the
subtraction performed in \eqref{Weff}, for nonequilibrium states.
Of course, \eqref{17} contains as a particular case the equilibrium
situations in which the subtracted term vanishes.

\subsection*{Excess work} 

Consider space-time dependent chemical potential and external field 
$(\lambda (t),E(t))$, $t\ge 0$,  
such that $(\lambda (t),E(t))$ converges to $(\lambda_1,E_1)$ 
as $t\to+\infty$ and an initial density profile $\rho$. 
We denote by $\bar\rho_1=\bar\rho_{\lambda_1,E_1}$ the stationary
profile associated to $(\lambda_1,E_1)$.

We introduce the \emph{excess work}
$W_\mathrm{ex} = W_\mathrm{ex}(\lambda,E,\rho)$ as the
difference between the renormalized energy 
$W^\textrm{ren}[\lambda,E,\rho]$ exchanged between the system and the
driving, 
and the minimal renormalized energy involved in a
quasi static transformation from $\rho$ to $\bar\rho_1$. Namely,
\begin{equation}
\label{17bis} 
\begin{split}
 W_\mathrm{ex}(\lambda, E,\rho) &= W^\textrm{ren}(\lambda,E,\rho)
- \min W^\textrm{ren}(\lambda', E', \rho)
\\
&= 
\int_0^\infty \!dt \int_{\Lambda} \!dx  \, J_\mathrm{S}(t,u(t)) 
\, \cdot \chi(u(t))^{-1} J_\mathrm{S}(t,u(t)) 
\end{split}
\end{equation}
where we used \eqref{16} and the minimum is take on all the paths
$(\lambda', E')$ such that 
$(\lambda'(+\infty), E'(+\infty))=(\lambda_1, E_1)$. 
In the case of transformations which are realized by a sequence of
equilibrium states, for each time $t$ the current $J(t)$ is purely
symmetric and the above definition coincides with \eqref{exwork}.

\subsection*{Relaxation path: excess work and  quasi potential}

Consider at time $t=0$ a stationary nonequilibrium profile $\bar\rho_0$
corresponding to some driving $(\lambda_0,E_0)$.
This system is put in contact with new reservoirs at chemical
potential $\lambda_1$ and a new external field $E_1$.
For $t> 0$ the system evolves according to the hydrodynamic equation
\eqref{2.1}--\eqref{2.3} with initial condition $\bar\rho_0$, time
independent boundary condition $\lambda_1$ and external field $E_1$. 
In particular, as $t\to \infty$ the system relaxes to $\bar\rho_1$.
Along such a path, in view of the orthogonality relation \eqref{ort},
the excess work is given by
\begin{equation*}
W_\mathrm{ex} (\lambda_1, E_1,\bar\rho_0) 
= \int_0^\infty \!dt \int_{\Lambda} \!dx  \, J (u(t)) 
\, \cdot \chi(u(t))^{-1} J_\mathrm{S}(u(t)) 
\end{equation*}
where $J_\mathrm{S}$ is computed by using the quasi potential
$V_{\lambda_1,E_1}$.  

By definition \eqref{jd} of the symmetric part of the current and by
an integration by parts, the previous expression is equal to
\begin{equation*}
\int_0^\infty \!dt \int_{\Lambda} \!dx  \, \nabla \cdot J (u(t)) 
\,  \frac{\delta V_{\lambda_1,E_1} (u(t)) }{\delta \rho} 
= -\, \int_0^\infty \!dt \int_{\Lambda} \!dx  \, \partial_t u(t) 
\,  \frac{\delta V_{\lambda_1,E_1}(u(t)) }{\delta \rho}.
\end{equation*}
We have therefore shown that
\begin{equation}
\label{u=qp}
W_\mathrm{ex} (\lambda_1, E_1,\bar\rho_0) = V_{\lambda_1,E_1} (\bar\rho_0)
- V_{\lambda_1,E_1} (\bar\rho_1) = V_{\lambda_1,E_1} (\bar\rho_0)
\end{equation}
which extends to nonequilibrium states the relation \eqref{exw=qp} between
the excess work and the quasi potential.

\subsection*{Time dependent transformations}

Instead of the transformations examined in the previous subsection,
where the external driving  is constant in time, we consider a
transformation with smooth space-time dependent chemical potential and
external field. We thus consider a path $(\lambda(t),E(t))$, $t\ge 0$,
such that $(\lambda(t),E(t))\to (\lambda_1,E_1)$ as $t\to +\infty$
fast enough. 
We denote by $\bar\rho_1=\bar\rho_{\lambda_1,E_1}$ the stationary
profile corresponding to $(\lambda_1,E_1)$ and let $(u(t),j(t))$,
$t\ge 0$ be the solution to the hydrodynamic equation
\eqref{2.1}--\eqref{2.3} with initial condition $u(0)=\rho$. Here
$\rho$ is an arbitrary density profile.

In this case, the computations which led to
\eqref{u=qp} give that the excess of work is equal to
\begin{equation*}
\begin{split}
W_\mathrm{ex} (\lambda , E, \rho) &=
-\, \int_0^\infty \!dt \int_{\Lambda} \!dx  
\,\frac{\delta V_{\lambda (t), E(t)}(u(t))}{\delta \rho} 
\, \partial_t u(t) 
\\
&=
- \int_0^\infty\!\!dt\: \frac{d}{dt}  V_{\lambda(t),E(t)}(u(t))
+ \int_0^\infty\!\!dt\: \big( {\partial_t} 
V_{\lambda(t),E(t)} \big)\,(u(t)) \;.
\\
&= V_{\lambda(0),E(0)}(\rho) \;+\; \int_0^\infty\!\!dt\: \big( {\partial_t} 
V_{\lambda(t),E(t)} \big)\,(u(t)) 
\end{split}
\end{equation*}
where we used that $u(t)\to\bar\rho_1$ as $t\to +\infty$ fast enough
and $V_{\lambda_1,E_1}(\bar\rho_1)=0$.

In particular, when we start from the stationary density profile
associated to $(\lambda(0),E(0))$, i.e.\
$\rho=\bar\rho_0=\bar\rho_{\lambda(0),E(0)}$, we conclude
\begin{equation}
\label{tbc}
W_\mathrm{ex} (\lambda, E,\bar\rho_0) 
= \int_0^\infty\!dt\, \big( {\partial_t} V_{\lambda(t),E(t)} \big)
(u(t)) \ge 0. 
\end{equation}
Note that the right hand side is not a total derivative and,
in particular, the excess work depends on the path of the driving
$(\lambda(t),E(t))$. From the previous formula we deduce that
excess work can be computed in terms of the time derivatives of the driving
forces. The inequality in \eqref{tbc}, which follows from \eqref{17bis}, is a
restatement of the Clausius inequality
\eqref{15}.

\subsection*{Quasi potential and specific relative entropy}

The relationship \eqref{ler} between the relative entropy and the
quasi potential extends, exactly with the same formulation, to
nonequilibrium states. We discuss only the case of stochastic lattice
gases.  Recall that $\Lambda\subset \bb R^d$ is the macroscopic
volume, and denote by $\Lambda_\epsilon$ the corresponding subset of the
lattice with spacing $\epsilon$, so that the number of sites in
$\Lambda_\epsilon$ is approximately $\epsilon^{-d} |\Lambda|$.  
Given the chemical potential $\lambda$ of the boundary reservoirs and
the external field $E$, let $\mu^{\lambda,E}_{\Lambda_\epsilon}$ be
the stationary measure of a driven stochastic lattice gas.  

Given $(\lambda_0,E_0)$ and $(\lambda_1,E_1)$, we claim that
\begin{equation}
\label{wp=rel}
\lim_{\epsilon\to 0} \epsilon^{d} 
\, S\big( \mu^{\lambda_0,E_0}_{\Lambda_\epsilon} \big|
\mu^{\lambda_1,E_1}_{\Lambda_\epsilon} \big) = \beta\, V_{\lambda_1,E_1}(\bar\rho_0),
\end{equation}
where $\beta=1/\kappa T$, the relative entropy $S$ has been defined
in \eqref{relen}, and $\bar\rho_0$ is the stationary profile
corresponding to $(\lambda_0,E_0)$.

We refer to Section~\ref{s:6.2} for a detailed derivation of
\eqref{wp=rel} under the assumptions that the stationary measures
satisfy a strong form of local equilibrium (that holds e.g.\ for the
boundary driven symmetric simple exclusion process). 
We next present a simple heuristic argument leading
to \eqref{wp=rel}. 
In view of the definition \eqref{relen} of the relative entropy we
have that
\begin{equation*}
\epsilon^d \, S\big( \mu^{\lambda_0,E_0}_{\Lambda_\epsilon} \big| 
\mu^{\lambda_1,E_1}_{\Lambda_\epsilon} \big) = 
\epsilon^{d} 
\sum_{\eta}  \mu^{\lambda_0,E_0}_{\Lambda_\epsilon}(\eta) \log \frac
{\mu^{\lambda_0,E_0}_{\Lambda_\epsilon}(\eta)} 
{\mu^{\lambda_1,E_1}_{\Lambda_\epsilon}(\eta)}.  
\end{equation*}
By the large deviation formula \eqref{ld}, we then get
\begin{equation*}
\begin{split}
\epsilon^d \, S\big( \mu^{\lambda_0,E_0}_{\Lambda_\epsilon} \big| 
\mu^{\lambda_1,E_1}_{\Lambda_\epsilon} \big) 
&\approx \epsilon^d \beta 
\sum_{\eta}  \mu^{\lambda_0,E_0}_{\Lambda_\epsilon}(\eta) \big[ V_{\lambda_1,E_1} 
(\rho_\epsilon(\eta)) -V_{\lambda_0,E_0} (\rho_\epsilon(\eta))\big]
\\
& \approx\beta 
\big[ V_{\lambda_1,E_1} (\bar\rho_0) -V_{\lambda_0,E_0} (\bar\rho_0)\big] 
 =\beta \, V_{\lambda_1,E_1} (\bar\rho_0)\;,
\end{split}
\end{equation*}
where $\rho_\epsilon(\eta)$ denotes the density profile associated to the
microscopic configuration $\eta$. In the final step we used the law of
large numbers for the microscopic density profile under the
probability $\mu^{\lambda_0,E_0}_{\Lambda_\epsilon}$.

Actually, the above argument is somewhat misleading.  The identity
\eqref{wp=rel} is not a consequence only of the large deviation
formula \eqref{ld}. It is in fact not difficult to construct
counterexamples to such a general statement.  Let, for instance,
$\mu^\beta_\epsilon$ be the Gibbs measure for a one-dimensional Ising
model at zero magnetic field and inverse temperature $\beta$ on a ring
with $\epsilon^{-1}$ sites.  
The magnetization satisfy the large deviation formula \eqref{ld} and
its typical value is zero for both ensembles so that the right hand
side of \eqref{wp=rel} vanishes. On the other hand, 
by a direct computation, for $\beta_0\neq \beta_1$, $\lim_{\epsilon}
\epsilon S( \mu^{\beta_0}_\epsilon| \mu^{\beta_1}_\epsilon) >0$.
Observe that this example does not contradict \eqref{wp=rel} as we are
comparing two ensembles in which we varied the temperature and not the
magnetic field. In this example, the correct formulation of
\eqref{wp=rel} would have been in terms of the large deviation
function for the energy, that is the extensive variable conjugated the
the intensive parameter that has been changed.

\section{Time dependent quasi potential}
\label{s:5}

In the previous section we have considered the case in which the
external driving changes over time scales that are comparable to or
longer than the typical relaxation times of the system. The
renormalized work has thus been defined by considering the values of 
the chemical potential and the external field frozen at a given time,
see \eqref{Weff}. In this section we consider a different approach,
suited for faster transformations, in which we take into account the
fact that the system has a finite relaxation time. 
We here define the renormalized work by using a time
dependent quasi potential which, at a given time, depends on the
previous history. 

Throughout all this section we fix a space-time dependent chemical
potential $\lambda(t)$ and an external field $E(t)$, where now
$-\infty<t<+\infty$. We assume that $(\lambda(t),E(t))$ converges
(fast enough) to $(\lambda_0,E_0)$ and $(\lambda_1,E_1)$ as $t\to
-\infty$ and $t\to +\infty$, respectively. We denote by $\bar\rho_0$
and $\bar\rho_1$ the stationary profiles corresponding to
$(\lambda_0,E_0)$ and $(\lambda_1,E_1)$, i.e.\
$\bar\rho_i=\bar\rho_{\lambda_i,E_i}$, $i=0,1$.

For $T_-<T_+$, denote by $I_{[T_-, T_+]}$ the action functional on the
set of paths $(u(t),j(t))$, $t\in [T_-,T_+]$,  defined as in \eqref{Idyn}
\begin{equation*}
I_{[T_-, T_+]}(u,j) = {\frac 14} 
\int_{T_-}^{T_+} \!dt \int_\Lambda \!dx \: 
\big[ j(t) - J(t,u(t)) \big] 
\cdot \chi(u(t))^{-1} \big[ j(t) - J(t,u(t)) \big] ,
\end{equation*}
where $(u(t),j(t))$ satisfy the continuity equation $\partial_t u
+\nabla \cdot j=0$ 
and  $J(t,\rho)$ is given in \eqref{2.2}. In particular, if
$(u(t),j(t))$ is a solution of the hydrodynamic equation \eqref{2.1}
then $I_{[T_-,T_+]}(u,j)=0$.

For two density profiles $\rho_-$, $\rho_+$, denote by 
$V_{\lambda,E}(T_-, \rho_-; T_+, \rho_+)$ the minimal action in the 
transition from $\rho_-$ to $\rho_+$ in the time interval $[T_-,T_+]$:
\begin{equation}
\label{tdvp}
V_{\lambda, E} (T_-, \rho_-; T_+, \rho_+)  =
\inf\,\big\{I_{[T_-,T_+]}(u,j)\,,\: u(T_-)=\rho_-\,,\: 
u(T_+)=\rho_+ \big\}.
\end{equation}
By a calculus of variations, similar to the one performed in classical
mechanics, $V_{\lambda, E}(T_-, \rho_-; t, \rho)$, as a function of
$t\in(T_-,T_+)$ and $\rho$, solves the time dependent Hamilton-Jacobi equation
\begin{equation}
\label{tdhj}
{\partial_t}\, V_{\lambda, E} + 
\int_\Lambda \!dx \:  
\nabla \frac{\delta V_{\lambda, E}}{\delta \rho} \, \cdot \, \chi(\rho) 
\, \nabla \frac{\delta V_{\lambda, E}}{\delta \rho} 
- \int_\Lambda \!dx \:  \frac{\delta V_{\lambda, E}}{\delta \rho} 
\, \nabla \cdot J(t,\rho) = 0 ,
\end{equation}
where ${\delta V_{\lambda, E}}/{\delta \rho}={\delta V_{\lambda, E}}
(T_-,\rho_-;t,\rho)/{\delta \rho}$ vanishes at the 
boundary $\partial \Lambda$ and $\rho$ satisfies the boundary
condition $f'(\rho(x)) = \lambda(t,x)$, $x\in\partial\Lambda$.

Let
\begin{equation*}
V_{\lambda, E}(\rho_-; t, \rho) = 
\lim_{T_-\to-\infty} V_{\lambda, E}(T_-, \rho_-; t, \rho).
\end{equation*}
By taking the limit $T_-\to -\infty$ the dependence on the initial
condition $\rho_-$ disappears so that
\begin{equation}
\label{tdqpsi}
V_{\lambda, E}(\rho_-; t, \rho) =   
V_{\lambda, E}(t, \rho) 
= \inf\,\big\{I_{(-\infty,t]}(u,j)\,,\: u(t)=\rho\,,\:
\lim_{s\to-\infty}u(s) =\bar\rho_0 \big\}.
\end{equation}
In fact, when $T_-\to-\infty$ the optimal path for the variational
problem on the right hand side of \eqref{tdvp} first essentially relaxes to 
$\bar\rho_0$ according to the hydrodynamic equation 
(since $(\lambda(s),E(s))\to (\lambda_0,E_0)$ as $s\to-\infty$) 
and then follows the optimal path for the right hand side
of \eqref{tdqpsi}.
Observe that $V_{\lambda, E}(t, \rho) $ is obtained by solving a time
dependent variational problem while the functional
$V_{\lambda(t),E(t)}(\rho)$ used in Section~\ref{s:4} is obtained
by solving a time independent variational problem with the chemical
potential and the external field frozen at time $t$.
We remark that if $\rho$ coincides with the solution of \eqref{2.1} at
time $t$ then $V_{\lambda,E}(t,\rho)=0$.

Note that $V_{\lambda, E}(t,\rho)$ provides a large deviation formula
analogous to \eqref{ld} in the case of time dependent chemical
potentials and external fields,
\begin{equation}
\label{ldtd}
\bb P^{\lambda,E} \big[ u_\epsilon (t) \approx \rho \big]
\asymp \exp\big\{ - \epsilon^{-d}\,\beta\, V_{\lambda, E}(t,\rho) \big\},
\end{equation}
where $\bb P^{\lambda,E}$ is the ensemble (defined on space-time
paths) corresponding to the time dependent chemical potential and
external field, $\epsilon$ is the scaling parameter, and
$u_\epsilon(t)$ is the empirical density at time $t$.
The asymptotics \eqref{ldtd} can be derived as follows. If we look at
the large deviations probability for a space-time path $(u(s),j(s))$,
$-\infty < s\le t$ of the empirical density and current we get 
\begin{equation}
\label{ldst}
\bb P^{\lambda,E} \big( (u_\epsilon (s), j_\epsilon(s)) \approx 
(u(s),j(s)),\, s\in (-\infty,t] \big)
\asymp \exp\big\{ - \epsilon^{-d}\beta I_{(-\infty,t]} (u,j) \big\}. 
\end{equation}
This formula has been derived in \cite{primo,curr} when the chemical
potential and external field do not depend on time. The argument can
be extended to the present setting.  By minimizing with respect to
the path $(u(s),j(s))$, $-\infty <s\le t$, with the constraint
$u(t)=\rho$ we deduce \eqref{ldtd}.

We observe that the functional $V_{\lambda, E}(t, \rho)$ still solves
the time dependent Hamil\-ton-Jacobi equation \eqref{tdhj}. Moreover, by
taking the limit $t\to \pm \infty$ we recover the time independent
quasi potentials associated to the chemical potentials and external
fields $(\lambda_0, E_0)$, $(\lambda_1, E_1)$, namely
\begin{equation*}
\lim_{t\to -\infty} V_{\lambda, E}(t, \rho) = V_{\lambda_0, E_0} (\rho), 
\quad 
\lim_{t\to +\infty} V_{\lambda, E}(t, \rho) = V_{\lambda_1, E_1}(\rho).
\end{equation*}

\subsection*{Renormalized work}

Let $V_{\lambda,E}(t,\rho)$ be the time dependent quasi potential
defined in \eqref{tdqpsi} in which we emphasize that
$V_{\lambda,E}(t,\rho)$ depends on the whole path $(\lambda(s), E(s))$
for $-\infty< s\le t$.
In analogy with \eqref{decur} we decompose the current as
\begin{equation}
\label{decurr2}
J(t,\rho) =  J_\mathrm{1}(t,\rho) 
+ J_\mathrm{2}(t,\rho),
\end{equation}
where 
\begin{equation}
\label{jd2}
J_\mathrm{1}(t,\rho)= -\chi(\rho) \nabla 
\frac {\delta V_{\lambda,E}(t,\rho)}{\delta\rho},
\end{equation}
and, recalling \eqref{2.2}, $J_\mathrm{2}(t,\rho)$ is defined via
\eqref{decurr2} by difference. 
Observe that the definition of $J_\mathrm{1}(t,\rho)$ 
differs from the symmetric current
$J_{\mathrm{S}}(t,\rho)$ introduced in \eqref{jd}.  
In fact in definition \eqref{jd} we introduced the ``thermodynamic
force'' $\delta V_{\lambda(t),E(t)}(\rho) /{\delta\rho}$ by
considering the quasi potential with the chemical potential and
external field frozen at time $t$ while in \eqref{jd2} we used the
time dependent quasi potential, i.e.\ we considered the
time dependent thermodynamic force $\delta V_{\lambda,E}(t,\rho) /{\delta\rho}$.
The difference among these two definition is the following. The
symmetric current $J_\mathrm{S}(t,\rho)$ in \eqref{jd} takes into
account only the values of the driving $\lambda, E$ at the time $t$
and not the actual state of the system, in particular it is
independent of the values $\lambda(s), E(s)$ for $s<t$. On the other
hand the current $J_\mathrm{1}(t,\rho)$ in \eqref{jd2} depends on the
actual state of the system and reflects the fact that the
system has a strictly positive relaxation time. 
Since $V(t,\rho)$ is minimal when $\rho$ coincides with the solution
of the hydrodynamic equation \eqref{2.1} at time $t$ we get that in
this case $J_1(t,\rho)=0$ or equivalently $J(t,\rho)=J_2(t,\rho)$.
In the quasi static limit, i.e.\ for transformations $\lambda,E$ which
vary very slowly, the definitions \eqref{jd} and \eqref{jd2} coincide.

In contrast with \eqref{ort}, the decomposition \eqref{decurr2} is not
orthogonal and the time dependent Hamilton-Jacobi equation
\eqref{tdhj} implies
\begin{equation}
\label{nonort}
\int_\Lambda\!dx\: J_\mathrm{1}(t,\rho) \cdot \chi(\rho)^{-1}
J_\mathrm{2}(t,\rho) =
\int_\Lambda\!dx\: 
\frac {\delta V_{\lambda,E}(t,\rho)}{\delta\rho} 
\, \nabla\cdot  J_\mathrm{2}(t,\rho) 
=  {\partial_t}\, V_{\lambda,E}(t,\rho).
\end{equation}

Fix a time window $[t,T]$ and let $(u(s),j(s))$, $ t\le s \le T$, 
be the solution of \eqref{2.1}--\eqref{2.3} with initial
condition  $u(t)=\rho$.  Here $\rho$ is an arbitrary density
profile (not necessarily the solution of the hydrodynamic equation at
time $t$). 
We now define the renormalized work
$\hat{W}^\textrm{ren}_{[t,T]}(\lambda,E,\rho)$ in the time interval
$[t,T]$ as
\begin{equation}
\label{Weff2}
\begin{split}
&\hat{W}^\textrm{ren}_{[t,T]} (\lambda,E,\rho) = 
\, W_{[t,T]} (\lambda,E,\rho) \\
&\quad
- \int_{t}^{T}\! ds \!\int_\Lambda \!dx\: J_\mathrm{2}(s,u(s)) \cdot
\chi(u(s))^{-1} J_\mathrm{2}(s,u(s))
- 2 \int_{t}^{T}\! ds \: {\partial_s} V_{\lambda,E}(s,u(s)).
\end{split}
\end{equation}
where $W_{[t,T]} (\lambda,E,\rho)$ is given in \eqref{W=}
and last term above takes into account the energy exchanged due to the
variation of the external driving in time.
By taking the limit $T\to +\infty$ and using \eqref{04} together with
\eqref{nonort} we deduce
\begin{equation}
\label{Weff2bis}
\begin{split}
\hat{W}^\textrm{ren}_{[t,+\infty)} (\lambda,E,\rho) = & 
\, F(\bar\rho_1)-F(\rho) 
\\
& + \int_{t}^{\infty}\! ds \!\int_\Lambda \!dx\: 
J_\mathrm{1}(s,u(s)) \cdot \chi(u(s))^{-1} J_\mathrm{1}(s,u(s)) 
\end{split}
\end{equation}
In particular, the renormalized work
$\hat{W}^\textrm{ren}_{[t,+\infty)}(\lambda,E,\rho)$  
satisfies  the Clausius inequality 
\begin{equation}
  \label{ci3}
  \hat{W}^\textrm{ren}_{[t,\infty)}(\lambda,E,\rho) 
  \ge F(\bar\rho_1)-F(\rho) =\Delta F
\end{equation}
where we recall that $\rho$ is the initial datum at time $t$ and
$\bar\rho_1=u(+\infty)$.  As we discuss below, definition
\eqref{Weff2} also yields the identity between the associated excess
work and the time dependent quasi potential $V(t,\rho)$.

\subsection*{Quasi static transformations}

The arguments of the previous sections concerning quasi static
transformations can be easily modified to the present setting.  Recall
that $\hat{W}^\textrm{ren}_{[t,+\infty)}(\lambda,E,\rho)$ involves the
current $J_1(s,\rho)$, for $t<s<+\infty$, as defined in \eqref{jd2},
which depends on the path $(\lambda(s),E(s))$ for $-\infty<s\le t$. As
the right hand side of \eqref{Weff2bis} depends also on $u(s)$ for $s>t$
which is determined by $(\lambda(s), E(s))$ for $t \le s<+\infty$, we
conclude that $\hat{W}^\textrm{ren}_{[t,+\infty)}(\lambda,E,\rho)$
depends on the whole path $(\lambda(s), E(s))$ for $-\infty <
s<+\infty$. 
In particular, a relevant statement of optimality of quasi static
transformation in the Clausius inequality \eqref{ci3} needs to include
the condition that the driving $(\lambda, E)$ is not changed in the time
interval $(-\infty,t)$.

As before we denote by $\rho$, which is an arbitrary density profile,
the initial datum of the density at time $t$. Given such time $t$ and
the density profile $\rho$, we claim that there exist a sequence of
smooth paths $(\lambda_\delta(s), E_\delta(s))$, $-\infty<s<+\infty$,
$\delta>0$ such that: (i) the history before time $t$ is not changed,
i.e.\ $(\lambda_\delta(s), E_\delta(s))=(\lambda (s), E (s))$ for
$-\infty<s<t$; (ii) at time $t+\delta$ the stationary profile
associated to $(\lambda_\delta,E_\delta)$ is $\rho$, i.e.\
$\bar\rho_{\lambda_\delta(t+\delta),E_\delta(t+\delta)} =\rho$; (iii)
the asymptotic state at time $s=+\infty$ is unchanged, i.e.\
$(\lambda_\delta(+\infty), E_\delta(+\infty)) = (\lambda_1,E_1)$; (iv)
in the quasi static limit $\delta\to 0$ equality in \eqref{ci3} is
achieved, i.e.\
\begin{equation*}
\lim_{\delta\to 0}\hat{W}^\textrm{ren}_{[t,\infty)}
(\lambda_\delta,E_\delta,\rho)
= F(\bar\rho_1)-F(\rho) =\Delta F.
\end{equation*}
The sequence $(\lambda_\delta(s), E_\delta(s))$, $-\infty<s<+\infty$,
$\delta>0$, can be constructed as in the previous sections and we omit
the details.

\subsection*{Excess work}
As before, given the time window $[t, +\infty)$ we let $(u(s),j(s))$, $s \in
[t,+\infty)$, be the solution of \eqref{2.1}--\eqref{2.3} with initial
condition  $u(t)=\rho$, where $\rho$ is an arbitrary density
profile. We then define the excess work along this path by 
\begin{equation}
  \label{Wex2}
  \begin{split}
    \hat{W}^\textrm{ex}_{[t,+\infty)} (\lambda,E,\rho) 
    & = \hat{W}^\textrm{ren}_{[t,+\infty)} (\lambda,E,\rho) 
    - \big[ F(\bar\rho_1) -F(\rho)\big] 
    \\
    &=
    \int_{t}^{\infty}\! ds \!\int_\Lambda \!dx\: J_\mathrm{1}(s,u(s)) \cdot
    \chi(u(s))^{-1} J_\mathrm{1}(s,u(s)).
  \end{split}
\end{equation}
We claim that
\begin{equation}
\label{hatwex}
  \hat{W}^\textrm{ex}_{[t,+\infty)}(\lambda,E,\rho) 
  = V_{\lambda,E}(t,\rho) -V_{\lambda,E}(+\infty,\bar\rho_1)
  = V_{\lambda,E}(t,\rho)
\end{equation}
where $V_{\lambda,E}(t,\rho)$ is the time dependent quasi potential.  
Observe that \emph{a priori} the excess work
$\hat{W}^\textrm{ex}_{[t,+\infty)}(\lambda,E,\rho)$, as it involves
the time integral of the current $ J_\mathrm{1}(s,u(s))$ on the time
window $[t,+\infty)$, should depend on the whole path
$(\lambda(s),E(s))$, $-\infty <s < +\infty$. However, 
the quasi potential $V_{\lambda,E}(t,\rho)$ on 
the right hand side of  \eqref{hatwex} depends only on the path 
$(\lambda(s),E(s))$ for $-\infty <s \le t$. 
Observe that if $\rho$ coincides with the
solution of the hydrodynamic equation \eqref{2.1} on the time interval
$(-\infty,t)$ evaluated at time $t$ then we get 
$\hat{W}^\textrm{ex}_{[t,+\infty)}(\lambda,E,\rho) =0$ as it is
apparent from the right hand side of \eqref{Wex2}. 

To prove \eqref{hatwex} we write $J_\mathrm{1}(t,u(t))$ as $J (t,u(t))
- J_\mathrm{2}(t,u(t))$. By using \eqref{jd2} and \eqref{nonort} we
deduce
\begin{equation*}
\begin{split}
 &\hat{W}^\textrm{ex}_{[t,+\infty)}(\lambda,E,\rho) 
= \int_t^\infty\!ds\!\int_\Lambda \!dx \:
J_\mathrm{1}(s,u(s)) \cdot \chi(u(s))^{-1} 
J_\mathrm{1}(s,u(s)) \\
&\qquad = \int_t^\infty\!ds\!\int_\Lambda \!dx \:
\frac {\delta V_{\lambda,E}(s,u(s))}{\delta\rho} 
\, \nabla \cdot J(s,u(s)) - \int_t^\infty\!ds 
\: \partial_s V_{\lambda,E} (s,u(s)).
\end{split}
\end{equation*}
Since $\nabla \cdot J(s,u(s)) = -\partial_s u(s)$, we get
\begin{equation*}
 \hat{W}^\textrm{ex}_{[t,+\infty)}(\lambda,E,\rho) 
=- \int_t^\infty\!ds \: \frac{d}{ds} V_{\lambda,E}(s,u(s)) 
= V_{\lambda,E}(t,\rho),
\end{equation*}
where we used that $V_{\lambda,E}(+\infty,u(+\infty)) 
= V_{\lambda_1,E_1}(\bar\rho_1)=0$.

\section{Stochastic lattice gases}
\label{s:6}

As basic microscopic model we consider a stochastic lattice gas in a
bounded domain with time dependent external field and boundary
conditions.
In the sequel we first exemplify the previous discussion in a simple
nonequilibrium model, the so-called zero range process, in which the
computations can be performed explicitly. We refer the reader e.g.\ to
\cite{rev} for a more general setting.
We then conclude this section by proving the relationship \eqref{wp=rel}
between the relative entropy and the quasi potential under
the assumption that the stationary ensemble satisfies a strong form of
local equilibrium.

\subsection{Time dependent zero range process}
\label{s:6.1}

Fix $\Lambda \subset \bb R^d$ and, given $\epsilon>0$, let
$\Lambda_\epsilon = (\epsilon^{-1} \Lambda) \cap \bb Z^d$ its discrete
approximation. The microscopic configuration is given by the
collection of occupation variables $\eta_i$, $i\in \Lambda_\epsilon$
so that $\eta_i$ is the number of particles at the site $i$.  The
dynamics can be informally described as follows.  At each site,
independently from the others, particles wait exponential times, whose
parameter depends only on the number of particles at that site, and
then jumps to a nearest neighboring site according to some transition
probability of a random walk on $\Lambda_\epsilon$.  Superimposed to
this bulk dynamics, to model the effect of the reservoirs, we have
creation and annihilation of particles, according to some birth and
death process, at the boundary of $\Lambda_\epsilon$.

\subsubsection*{Microscopic dynamics}
To define formally the microscopic dynamics, recall that a continuous
time Markov chain $\eta(\tau)$ on some state space $\Omega$ can be
described in term of its time dependent infinitesimal \emph{generator}
$L_\tau$ defined as follows. Let $f\colon\Omega\to \bb R$ be an observable,
then
\begin{equation}
\bb E \big( f (\eta ({\tau+h})) \big| \eta(\tau) \big) = (L_\tau
f)(\eta(\tau)) \, h  + o(h)
\end{equation}
where $\bb E(\,|\,)$ is the conditional expectation, so that the
\emph{expected} infinitesimal increment of $f(\eta(\tau))$ is
$(L_\tau f)(\eta(\tau))\, d\tau$. The transition probability of the Markov
process $\eta(\tau)$ is then given by the kernel of the semigroup
generated by $L_\tau$.

For the zero range process with time depend external field $E=E(t,x)$ and
chemical potential $\lambda=\lambda(t,x)$ (where $t$ and $x$ are the
macroscopic time and space variables), the generator $L_\tau$ is given by 
\begin{equation*}
  L_\tau = L_{\tau,0} + L_{\tau,1}
\end{equation*}
where $L_{\tau,0}$ describes the bulk dynamics and $L_{\tau,1}$ the boundary
dynamics; they are given by
\begin{equation}
\label{GEN} 
\begin{array}{lll}
{\displaystyle
L_{\tau,0} f(\eta)} &=& {\displaystyle 
\sum_{\substack{i,j\in\Lambda_\epsilon \\ |i-j|=1}}
g(\eta_i) \,e^{  (1/2) \epsilon \: E( \epsilon^2 \tau, \epsilon
  (i+j)/2 ) \cdot (j-i)} 
\big[ f(\eta^{i,j}) -f (\eta)\big] 
}
\\
{\displaystyle
L_{\tau,1} f(\eta)} &=& {\displaystyle
\sum_{\substack{i\in\Lambda_\epsilon, j \not\in\Lambda_\epsilon 
\\ |i-j|=1}} \Big\{
g(\eta_i) 
\,e^{  (1/2) {\epsilon} \:  
E(\epsilon^2 \tau, \epsilon (i+j)/2) \cdot (j-i)} 
\big[ f(\eta^{i,-}) -f (\eta)\big]
}
\\
&&\quad
{\displaystyle
+ e^{ \lambda(\epsilon^2 \tau,\epsilon j) + (1/2) \epsilon \: 
  E(\epsilon^2 t, \epsilon (i+j)/2) \cdot (i-j)} 
  \big[f(\eta^{i,+}) - f(\eta)\big]  \Big\} 
}
\end{array}
\end{equation}
in which 
\begin{equation}
\label{exy}
\eta^{i,j}_k  = \left\{
\begin{array}{ccl}
\eta_k &\hbox{if}& k\neq i,j \\
\eta_k -1 &\hbox{if}& k=i \\
\eta_k +1  &\hbox{if}& k=j 
\end{array}
\right.
\end{equation}
is the configuration obtained from $\eta$ when a
particle jumps from $i$ to $j$, and
\begin{equation}
\label{expm}
\eta^{i,\pm}_k = 
\left\{
\begin{array}{ccl}
\eta_k &\hbox{if}& k\neq i \\
\eta_k \pm 1 &\hbox{if}& k=i 
\end{array}
\right.
\end{equation}
is the configuration where we added (respectively subtracted) one particle at
$i$. 

The function $g$ describes the jump rate. More precisely, if at some
site there are $k\ge 1$ particles, each one independently waits an
exponential time with parameter proportional to $g(k)/k$ and then
jumps to one of the neighboring sites with a transition probability
which depends on the external field $E$. We also set $g(0)=0$ so that
no jumps occur when the site is empty.  In the special case $g(k)=k$
the dynamics introduced above represents the evolution of the
occupation numbers $\eta_i$ for not interacting random walks in the
space-time dependent external field $E$ on $\Lambda_\epsilon$ with the
appropriate boundary conditions depending on $\lambda$.  For
simplicity of notation, we did not introduced the dependence on the
temperature in the model.

Denoting by $\mu_\epsilon(\tau, \,\cdot\,)$ the distribution of
the occupation variables $\eta_i$, $i\in\Lambda_\epsilon$ at time
$\tau$, then it satisfies
\begin{equation}
\label{muinvpt}
\sum_{\eta} \mu_\epsilon(\tau_1,\eta) \, p_{\tau_1,\tau_2} (\eta,\eta') =
\mu_\epsilon (\tau_2,\eta')\;,\quad \tau_1\le \tau_2\;, 
\end{equation}
where $p_{\tau_1,\tau_2} (\eta,\eta')$ is  the transition probability
associated to the generator $L_\tau$, i.e., the kernel of the operator
\begin{equation}
  \label{semigroup}
  P_{\tau_1,\tau_2} = \mc T \exp \Big\{ \int_{\tau_1}^{\tau_2}\!d\tau
  \, L_\tau \Big\} 
\end{equation}
where $\mc T$ denotes the time ordering.

\subsubsection*{Invariant measure}

We consider here the case in which the driving $(\lambda, E)$ does not
depend on time, so that the semigroup $P_{\tau_1,\tau_2}$ in
\eqref{semigroup} depends only on $\tau_2-\tau_1$ and is given by
$P_{\tau_2-\tau_1}=\exp\big\{(\tau_2-\tau_1)L\big\}$ where $L$ is the
time independent generator. In this case we next discuss the invariant
measure of the microscopic dynamics.

Since the Markov chain is irreducible (it is possible to go with
positive probability from any configuration to any other), under very
general hypotheses on the function $g(k)$ there exists a unique
invariant measure. This is the time independent probability
$\mu_\epsilon$ on the 
configuration state which solves \eqref{muinvpt}. 
It is remarkable that such invariant measure can be constructed
explicitly and it is product, see \cite{DF} for the one dimensional case.

Fix a time independent chemical potential $\lambda$ and external field $E$.
Let $\phi_\epsilon(i)$, $i\in\Lambda_\epsilon$, be the solution of the equations
\begin{equation}
\label{laminv}
\begin{cases}
\displaystyle{ \sum_{j \sim i}
\Big[ \phi_\epsilon(j) \,e^{  (1/2) \epsilon \: E(\epsilon (i+j)/2) \cdot (i-j)} 
- \phi_\epsilon(i) \,e^{  (1/2) \epsilon \: E( \epsilon (i+j)/2) \cdot (j-i)} 
\Big]
= 0}, 
& i \in\Lambda_\epsilon 
\\
\phi_\epsilon(i) = \exp\big\{ \lambda(\epsilon i)\big\}, 
& i\not\in\Lambda_\epsilon 
\end{cases}
\end{equation}
where the sum runs over the nearest neighbors of $i$.
The invariant measure $\mu_\epsilon$ is the product measure
$\mu_\epsilon=\prod_{i\in\Lambda_\epsilon} \mu_{\epsilon,i}$ obtained by
taking the product of the marginal distributions
\begin{equation}
\mu_{\epsilon,i}
(\eta_i = k) = \frac {1}{Z(\phi_\epsilon(i))} \; 
{\frac {\phi_\epsilon(i)^k}{g(1)\cdots g(k)}} 
\label{INV}
\end{equation}
where 
\begin{equation}
\label{Z=}
Z(\varphi) = 1 + \sum_{k=1}^{\infty}{\frac {\varphi^k}{g(1)
\cdots g(k)}}
\end{equation}
is the normalization constant.
The fact that $\mu_\epsilon$ is the invariant measure can be verified by
showing that $\sum_{\eta} \mu_\epsilon(\eta) L f (\eta) = 0$ for any observable $f$.  

Consider now an homogeneous equilibrium state which is obtained by
choosing $E=0$ and $\lambda$ constant.  
In this case $\phi_\epsilon=\exp\{\lambda\}$ so that the invariant
measure is Gibbs with Hamiltonian
\begin{equation*}
  H_\epsilon(\eta) = \sum_{i\in\Lambda_\epsilon} \sum_{k=1}^{\eta_i}  \log g(k)
\end{equation*}
where, comparing with \eqref{gibbs}, we understand that $\beta=1$.
In particular, in the stationary ensemble there is no interaction among
particles on different sites.

The computation of the pressure, see \eqref{press}, can be done
explicitly and by Legendre duality one obtains that the specific free
energy is given by
\begin{equation}
  \label{fezr}
  f(\rho) = \rho \log \Phi(\rho)  - \log Z(\Phi(\rho))
\end{equation}
where $\Phi\colon \bb R_+ \to \bb R_+$ is the inverse of the
strictly increasing function $R(\varphi)= \varphi\,
Z'(\varphi)/Z(\varphi)$.

\subsubsection*{Hydrodynamic limit}

For $x\in\Lambda$, $t\ge 0$, we introduce the \emph{empirical density} as
\begin{equation}
u_\epsilon(t,x) = \epsilon^d \sum_{i\in\Lambda_\epsilon} 
\eta_i(\epsilon^{-2} t)  \: \delta \big(x - \epsilon i \big)
\label{ED}
\end{equation}
where $\delta$ denotes the Dirac function. Given $B\subset
\Lambda$ let $B_\epsilon= \epsilon^{-1}B \cap \bb Z^d$. Then  
\begin{equation*}
\int_B \!dx\: u_\epsilon(t,x) = \epsilon^d 
\sum_{i\in B_\epsilon}  \eta_i( \epsilon^{-2} t) 
\end{equation*}
is the total mass in the volume $B$ at the macroscopic time $t$.

It is not difficult to extend the standard arguments of hydrodynamic
limits, see e.g.\ \cite{primo} for a heuristic derivation and
\cite{KL,S} for a rigorous analysis, to the present time dependent
setting. The formal statement is that in the scaling limit $\epsilon\to
0$ a law of large numbers for the empirical density holds. More
precisely, if at time $t=0$ the empirical density converges to some
profile $\rho$ (i.e.\ $u_\epsilon(0,x)\to \rho(x)$, $x\in\Lambda$)
then at time $t$ the empirical density $u_\epsilon(t)$ converges to
the solution $u(t)$ of the hydrodynamic equation
\begin{equation}
  \label{hezr}
  \begin{cases}
    \partial_t u + \nabla \cdot \big( \Phi(u) E(t) \big)
    = \Delta \Phi (u),& (t,x) \in \bb R_+\times \Lambda 
    \\
    \Phi \big( u(t,x) \big) = 
    \exp\big\{ \lambda(t,x) \big\},
      & (t,x) \in \bb R_+\times \partial\Lambda \\
    u(0,x) =\rho(x), & x\in\Lambda 
  \end{cases}
\end{equation}
where $\Delta$ is the Laplacian and the function $\Phi\colon
\bb R_+\to \bb R_+$ has been introduced above.  In particular, by
comparing \eqref{hezr} with \eqref{2.1}--\eqref{2.2} for the zero
range process the diffusion coefficient is $D= \Phi'$ and the mobility
is $\chi=\Phi$. As follows from \eqref{fezr} the local Einstein
relation \eqref{ein_rel} holds.  Finally, since $f'(\rho)=\log
\Phi(\rho)$ (also this follows from \eqref{fezr}) the boundary conditions
above agree with \eqref{2.3}.

The fluctuation formula \eqref{ldst} with the functional $I$ given by
\eqref{Idyn} is discussed in \cite{primo,curr} for time-independent
driving. The arguments can be extended to the present time-dependent
setting.

\subsubsection*{Microscopic work}

We next present the microscopic definition of the work done by
external field and the boundary reservoirs. 
To this aim, we first recall the definition of the \emph{empirical
  current}, see e.g., \cite{curr}.  Fix a path $\eta(\tau)$ of the
microscopic configuration.  Given an oriented bond $(i,j)$, let
$\mc{N}_{i,j}(\tau)$ be the number of particles that jumped from $i$
to $j$ in the time interval $[0,\tau]$.  Here we adopt the convention
that $\mc{N}_{i,j}(\tau)$ is the number of particles created at $j$
due to the reservoir at $i$ if $i\not\in\Lambda_\epsilon$, $j
\in\Lambda_\epsilon$, and that $\mc{N}_{i,j}(\tau)$ is the number of
particles that left the system at $i$ by jumping to $j$ if
$i\in\Lambda_\epsilon$, $j \not\in\Lambda_\epsilon$. The difference
$\mc{J}_{i,j}(\tau)= \mc{N}_{i,j}(\tau) - \mc{N}_{j,i}(\tau)$ is the
net number of particles flown across the bond $(i,j)$
in the time interval $[0,\tau]$.  The instantaneous current across
$(i,j)$, denoted by $J_{i,j}$, is defined as $J_{i,j}= d
\mc{J}_{i,j}/d\tau$.  Of course, $J_{i,j}$ is a sum of
$\delta$-functions localized at the jump times with weight $+1$,
respectively $-1$, if a particle jumped from $i$ to $j$, respectively
from $j$ to $i$.

Let now $(\lambda(\cdot),E(\cdot))$  be a path of the external
driving and denote by $\eta(\tau)$ the corresponding microscopic
trajectory. The natural microscopic definition of the work exchanged
between the system and the external driving in the time interval
$[0,\tau]$ is 
\begin{equation}
\label{micwork}
\begin{split}
\mc W_{[0,\tau]}  =  &- \sum_{i\in\Lambda_\epsilon,j \not\in\Lambda_\epsilon}
 \int_0^\tau\!d\tau'\, \lambda(\epsilon^2 \tau',\epsilon j) J_{i,j}(\tau')
\\
& + \frac 12 \sum_{(i,j)}   \int_0^\tau\!d\tau'\,   \epsilon 
E\big(\epsilon^2 \tau',\epsilon \,\tfrac{i+j}2\big) \cdot (j-i) \, J_{i,j}(\tau')
\end{split}
\end{equation}
where the second sum is carried out over all bonds intersecting $\Lambda_\epsilon$. 
We emphasize that the above definition is given in terms of
microscopic quantities, indeed the dependence on the scaling parameter $\epsilon$ is due to the
fact that we have considered the external field of order $\epsilon$,
see \eqref{GEN}, and the drivings as functions of the macroscopic variables.

We now consider the scaling limit of the microscopic work. We thus set
$\tau=\epsilon^{-2}T$ and assume that the initial configuration of
particles corresponds to a density profile $\rho$. 
In view of the law of large numbers for the empirical current, see
e.g., \cite{curr}, 
as $\epsilon \to 0$
\begin{equation}
  \epsilon^{d}  \, \mc W_{[0,\epsilon^{-2} T]} \longrightarrow W_{[0,T]}
\end{equation}
where the right hand side is the macroscopic work defined in
\eqref{04}.
The fluctuations properties of $\mc W$ as $\epsilon\to 0$ can be
derived from those of the empirical current \cite{curr}.

\subsubsection*{Quasi potential}

We discuss first the case of time-independent driving.
Since for the zero range process the invariant measure is product, the
fluctuation formula \eqref{ld} can be proven directly. By
straightforward computations, see \cite{primo} for the case $E=0$, we
get that the quasi potential $V_{\lambda,E}$ is given by 
\begin{equation}
  \label{qpzr}
  V_{\lambda,E}(\rho) = \int_\Lambda\!dx\:
  \Big[ \rho \log \frac{\Phi(\rho)}{\bar\phi}  
  - \log \frac{Z(\Phi(\rho))}{Z(\bar\phi)} \Big]
\end{equation}
where $\bar\phi = \Phi(\bar\rho_{\lambda,E})$ in which
$\bar\rho_{\lambda,E}$ is the stationary solution of \eqref{hezr}.  It
is also simple to check that the function $\phi_\epsilon$ which solves
\eqref{laminv} converges to $\bar\phi$ in the scaling limit
$\epsilon\to 0$. In terms of the macroscopic fluctuation theory, a
couple of integration by parts show that the right hand side of
\eqref{qpzr} is a stationary solution of the Hamilton-Jacobi equation
\eqref{tdhj} and this provides an alternative proof of the fluctuation
formula \eqref{ld}.

By using the explicit formula \eqref{INV} for the invariant measure
together with the convergence of $\phi_\epsilon$ to $\bar\phi$
computations analogous the ones presented in the Gibbsian setting show
that the relationship \eqref{wp=rel} between the limiting relative
entropy and the quasi potential holds.

Since in this case the quasi potential has an explicit expression the
decomposition \eqref{decur} of the current $J(\rho)$ is 
\begin{equation}
  \label{dczr}
  \begin{split}
  &J_\mathrm{S}(\rho) = - \Phi(\rho) \big[ 
  \nabla \log \Phi(\rho) -\nabla \log \bar\phi\big] 
  \\
  &J_\mathrm{A}(\rho) = \Phi(\rho) \big[ E- \nabla \log\bar\phi \big] 
  \end{split}
\end{equation}
where we recall that $\bar\phi=\Phi(\bar\rho_{\lambda,E})$.  
In particular, the dependence on $(\lambda,E)$ in $J_\mathrm{S}$
appears only through the stationary solution $\bar\rho_{\lambda,E}$. 
This is a special feature of the zero range process. 

According to the discussion in Section~\ref{s:4}, the power needed to
maintain the zero range process in a nonequilibrium stationary state
is 
\begin{equation*}
  \int_\Lambda\!dx \: \bar\phi\,
  \big[ E- \nabla \log\bar\phi \big]^2. 
\end{equation*}

\subsubsection*{Time dependent quasi potential}

The time dependent Hamilton-Jacobi equation \eqref{tdhj} has not a
simple solution in general. However, when $\Phi(\rho)=\rho$, that
corresponds to the case of independent random walks, it holds 
\begin{equation}
  \label{tdqpzr}
  V_{\lambda,E}(t,\rho)= \int_\Lambda\!dx\:
  \Big[ \rho \big( \log \frac{\rho}{\psi(t)} -1\big)  
  +\psi(t) \Big]
\end{equation}
where $\psi(t)=\psi(t,x)$ is obtained as the value at time $t$ of the
solution to 
\begin{equation}
\label{heat}
  \begin{cases}
    \partial_s \psi +\nabla\cdot\big( \psi \,E(s) \big) = \Delta \psi,  
    & (s,x)\in (-\infty,t)\times \Lambda \\
    \psi(s,x) = \exp\big\{\lambda(s,x)\big\}, 
    & (s,x)\in (-\infty,t)\times \partial\Lambda \\
    \lim_{s\to -\infty} \psi(s) = \bar\rho_0
  \end{cases}
\end{equation}
where $\bar\rho_0$ is the density profile at time $-\infty$.
From the above expression is apparent that $V_{\lambda,E}(t,\cdot)$
depends on the path  $(\lambda(s),E(s))$ for $s\in (-\infty,t]$. 
On the other hand the quasi potential $V_{\lambda(t),E(t)}$ with the
values of $\lambda,E$ frozen at time $t$ is obtained by replacing
$\psi(t)$ in \eqref{tdqpzr} with the solution of  
$\nabla\cdot \big(\psi \, E(t) \big) =\Delta\psi$ with the boundary
condition $\psi(x) = \exp\big\{\lambda(t,x)\big\}$,
$x\in\partial\Lambda$. That is by replacing $\psi$ with
$\bar\rho_{\lambda(t),E(t)}$. 
The proof of the representation \eqref{tdqpzr} amounts to
straightforward computations and it is omitted.

\subsection{Relative entropy between nonequilibrium stationary states}
\label{s:6.2}

We next give some mathematical details on the relationship
\eqref{wp=rel} which expresses the relative entropy between two nonequilibrium
states in terms of  the quasi potential. 
We first present a general argument
which shows, without any further assumption, that an inequality is always
satisfied. We then show that equality holds if the stationary
ensembles satisfies a strong form of local equilibrium. As proven in \cite{BL} 
this condition holds for the boundary driven symmetric simple
exclusion process. 
We also remark that for this model the validity of \eqref{wp=rel} has
been already proven (with a different motivation) in \cite{Ba,DLS} 
in the particular case in which the reference ensemble is an equilibrium state.

Recall the definition \eqref{relen} of the relative entropy
$S(\nu|\mu)$ of the probability $\nu$ with respect to $\mu$. 
Consider two sequences of probabilities $\nu_n$ and $\mu_n$ on
some space $E$. We assume  that $\mu_n$ satisfies the large deviation formula
\begin{equation}
\label{ld2}
  \mu_n \big( \mathcal O_x \big) \asymp \exp\big\{ - n \, V(x) \big\}
\end{equation}
where $ \mathcal O_x$ is a small neighborhood of $x$ and the 
\emph{rate function} $V$ is a function on $E$. We also assume that $\nu_n$ satisfies the
law of large numbers 
\begin{equation*}
  \nu_n \big( \mathcal O_{\bar x}^\mathrm{c} \big) \longrightarrow 0
\end{equation*}
where $\bar x \in E$ and $ \mathcal O_{\bar x}^\mathrm{c}$ denotes the
complementary set of $\mathcal O_{\bar x}$. We then claim that the inequality 
\begin{equation}
  \label{relenin}
  \varliminf_{n\to +\infty} \frac 1n \, S(\nu_n|\mu_n)  \ge V(\bar x) 
\end{equation}
holds.  Indeed, recall the variational representation of the relative
entropy, see e.g.\ \cite[Appendix A1]{KL}
\begin{equation*}
S(\nu_n| \mu_n )= \sup_f \Big\{ \int \!d\nu_{n}\: f 
-  \log \int \!d\mu_{n}  \: e^{f} \Big\}
\end{equation*}
where the supremum is carried out over the functions $f$ on $E$. 
By choosing $f$ equal to $n\, V$ we get 
\begin{equation}
\label{prein}
\frac 1n  S(\nu_n| \mu_n ) \ge  \int \! d\nu_n \: V   - 
\frac 1{n} \log \int\! d\mu_{n}  \: e^{n V}.  
\end{equation}
Recall the Laplace-Varadhan theorem, see e.g.\ \cite[Theorem
4.3.1]{DZ}, which states that - under the assumption \eqref{ld2} - for
each function $\phi$ on $E$ it holds
\begin{equation*}
  \lim_{n\to\infty} \frac 1{n} \log \int\! d\mu_{n}  \: e^{n \phi} = \sup_{x\in E}
  \big\{ \phi(x) - V(x) \big\}.
\end{equation*}
In view of the the law of large numbers for $\nu_n$, the inequality
\eqref{relenin} now follows from \eqref{prein}.

\medskip 
Consider a stochastic lattice gas in the domain $\Lambda_\epsilon =
(\epsilon^{-1}\Lambda) \cap \bb Z^d$. Assume for simplicity that the
external field $E$ vanishes.  Given a time independent chemical
potential $\lambda(x)$ we denote by $\mu_\epsilon^{\lambda}$ the
associated stationary ensemble. Observe that $\mu_\epsilon^{\lambda}$
is a probability on $\bb N^{\Lambda_\epsilon}$. As for the zero range
process, we denote by $\eta_i=0,1,\ldots$ the number of particles at
the site $i\in\Lambda_\epsilon$.

Given $\delta>0$, we decompose the domain $\Lambda_\epsilon$ into 
small boxes $B_1, B_2, \ldots$ of size $\delta \epsilon^{-1}$ and
denote  by $\mb N=(N_1, N_2, \ldots )$ the number of particles in each box. 
We let $\nu_\epsilon^{\lambda}(\mb N)$ the probability of having
$N_1$ particles in the box $B_1$, $N_2$ particles in the box $B_2$,
and so on. Namely,
\begin{equation*}
\nu_\epsilon^{\lambda}(\mb N) 
= \mu_\epsilon^{\lambda}\Big(  
\sum_{i \in B_1} \eta_i = N_1,\: 
\sum_{i \in B_2} \eta_i = N_2, \ldots \Big). 
\end{equation*}
We also introduce the \emph{conditional ensemble}, denoted by
$\mu_\epsilon^{\lambda}(\cdot| \mb N)$, as the probability
$\mu_\epsilon^{\lambda}$ conditioned to have $N_1$ particles in the
box $B_1$, $N_2$ particles in the box $B_2$, and so on.

For a box $B\subset \bb Z^d$, $n\ge 0$, denote by
$\mu^{\text{can}}_{B,n}$ the equilibrium canonical measure on $B$ with
$n$ particles, that is
\begin{equation*}
\mu^{\text{can}}_{B,n} (\eta) \;\propto\; \exp\big\{ - \beta H_B(\eta)\big\}
\end{equation*}
where $H_B(\eta)$ is the energy of a configuration $\eta$ with $n$
particles in $B$.

We shall assume that the conditional ensemble $\mu_\epsilon^{\lambda}
(\cdot| \mb N)$ is close to the product of the canonical ensembles:
\begin{equation*}
\mu_\epsilon^{\lambda} (\eta| \mb N) \approx  
\prod_\ell \mu^{\text{can}}_{B_\ell,N_\ell} (\eta) \; .
\end{equation*}
in the sense that
\begin{equation}
\label{00}
\epsilon^d \, 
\log  \frac{ \mu_\epsilon^{\lambda} (\eta| \mb N)}
{\prod_\ell \mu^{\text{can}}_{B_\ell,N_\ell} (\eta)} \;\to\; 0
\end{equation}
uniformly over $\eta$ as we let first $\epsilon \to 0$ and then
$\delta\to 0$.  As proven in \cite{BL}, this condition is satisfied
for the boundary driven one-dimensional symmetric simple exclusion
process.

We prove the equality \eqref{wp=rel} under the previous assumption.  Fix
two chemical potentials $\lambda_0$, $\lambda_1$.  By definition of
the relative entropy \eqref{relen},
\begin{equation*}
\epsilon^d S(\mu_\epsilon^{\lambda_0} | \mu_\epsilon^{\lambda_1} )
\;=\; \epsilon^d \sum_{\eta} \mu_\epsilon^{\lambda_0} (\eta)
\log \mu_\epsilon^{\lambda_0} (\eta) \;-\; \epsilon^d \sum_{\eta} 
\mu_\epsilon^{\lambda_0} (\eta) \log \mu_\epsilon^{\lambda_1}
(\eta)\;. 
\end{equation*}
Rewrite the expressions inside the logarithms as
$\mu_\epsilon^{\lambda_i} (\eta| \mb N) \nu_\epsilon^{\lambda_i}(\mb
N)$. By \eqref{00}, the contribution to the sum of the term $\log
\{\mu_\epsilon^{\lambda_0} (\eta| \mb N)/ \mu_\epsilon^{\lambda_1}
(\eta| \mb N)\}$ vanishes as $\epsilon\to 0$ and then $\delta\to 0$.
It remains to estimate the limit
\begin{equation*}
\epsilon^d \sum_{\mb N} \nu_\epsilon^{\lambda_0}(\mb N)
\log \nu_\epsilon^{\lambda_0}(\mb N) \;-\;
\epsilon^d \sum_{\mb N} \nu_\epsilon^{\lambda_0}(\mb N)
\log \nu_\epsilon^{\lambda_1}(\mb N) \;.
\end{equation*}
in view of the law of large numbers, we expect
$\nu_\epsilon^{\lambda_0}$ to concentrate on the density profile
$\bar\rho_0$, while $\log \nu_\epsilon^{\lambda_i}(\mb N)$ converges
to $- \beta V_{\lambda_i} (\bar\rho_0)$. This statement concludes the
proof of \eqref{wp=rel}. 
For boundary driven symmetric simple exclusion processes, by adapting
the arguments in \cite{BL}, also the last step can be justified
rigorously.

\section{Langevin dynamics}
\label{sec4}

We next illustrate the general thermodynamic theory in the simpler
context of Langevin dynamics which is nowadays very popular, see e.g.\
\cite{Se} for a recent review. 
In the Smoluchowski approximation, the motion of a particle in a
viscous $d$-dimensional medium is described by the Langevin equation
\begin{equation}
\label{22}
\gamma \dot X_t \;=\; - \nabla U (X_t) 
\;+\; q\, E(t,X_t) \;+\; \sqrt{\frac {2 \gamma}\beta} \, \dot W_t,
\end{equation}
where $U$ is the reference potential, $\beta = 1/\kappa T$, $\gamma$
is the friction coefficient, $q$ is the charge, $E(t,x)$ is an applied
field, e.g. an electric field, and $W_t$ is a $d$-dimensional Brownian
motion.  We discuss the zero temperature limit $\beta\to\infty$ which
is analogous to the thermodynamic limit for stochastic lattice
gases, 

In the limit $\beta\to \infty$ the Smoluchowski equation becomes the
deterministic equation
\begin{equation}
\label{23}
\gamma \, \dot x_t \;=\; - \nabla U (x_t) \;+\; q\, E(t,x_t).
\end{equation}
Fix an initial state $\bar x_0$, a path $E(t,\cdot)$, $0\le t\le T$,
and denote by $x_t$ the solution of \eqref{23} with initial condition
$\bar x_0$.  The work done by the applied field $E$ in the time
interval $[0, T]$, denoted by $W_{[0, T]} (\bar x_0, E)$, is given by
\begin{equation}
\label{25}
W_{[0, T]}(\bar x_0, E) \;=\; q
\int_{0}^{T} dt \, E(t,x_t) \cdot \dot x_t .
\end{equation}

\subsubsection*{Excess work and quasi potential}

Assume that $E$ is a time independent gradient, $E= -\nabla \Phi$, which
corresponds to the case of equilibrium states.  Given a time independent
potential $\Phi$, denote by $\bar x_\Phi$ the minimum point of
$U+q \Phi$, assumed to be unique and a global attractor for the flow
\eqref{23}.  Fix two time independent potentials $\Phi_0$, $\Phi_1$.
Consider a particle initially at the position $\bar x_0 = \bar
x_{\Phi_0}$ which is driven to a new position $\bar x_1 = \bar
x_{\Phi_1}$ by changing the potential in time in a way that
$\Phi(t)=\Phi_0$ for $t\le 0$ and $\Phi(t)=\Phi_1$ for $t\ge T$, where
$T$ is some fixed positive time.  Let $x(t)$, $t \ge 0$, be the
solution of \eqref{23} with initial condition $\bar x_0$.  Since the
potential is equal to $\Phi_1$ for $t\ge T$, it holds $x(t)\to\bar
x_1$ as $t\to+\infty$. Moreover, as $\bar x_1$ is an equilibrium
state, $x(t)$, $\dot x(t)$ relax exponentially fast to $\bar x_1$,
$0$, respectively. The integral in \eqref{25} is thus convergent for
$T\to\infty$ and we deduce
\begin{equation}
\label{20}
\begin{split}
W(\bar x_0,E) &= W_{[0,\infty)}(\bar x_0, E) =  
\int_{0}^{\infty} \!dt \, 
\big[ \gamma \, \dot x_t + \nabla U (x_t) \big] \cdot \dot x_t  
\\
& = U(\bar x_1) - U(\bar x_0) \;+\; \gamma
\int_{0}^{\infty} dt  \, |\dot x_t|^2 
\\ 
& \ge \Delta U ,
\end{split}
\end{equation}
which expresses the Clausius inequality in this setting.  By arguing
as in Section~\ref{s:es}, we can show that in the quasi static limit,
obtained by letting $\Phi(t)$ change in time very slowly, the equality 
$W(\bar x_0, E) = U(\bar x_1) - U(\bar x_0)$ holds.

Define the \emph{excess} work, $W_{\rm ex} (\bar x_0, E)$, as the
difference between the work performed by the applied field and the
work involved in a quasi static transformation from $\bar x_0$ to
$\bar x_1$, namely
\begin{equation*}
W_\mathrm{ex} = W(\bar x_0, E)  - \min W \;=\; \gamma
\int_0^\infty \!dt \, |\dot x_t|^2.
\end{equation*}

Consider the equilibrium point $\bar x_0$ associated to a potential
$\Phi_0$ and the path $\gamma \dot x_t \;=\; - \nabla U (x_t) \;-\;
q\, \nabla \Phi_1(x_t)$, for some $\Phi_1\not = \Phi_0$, with initial
condition $\bar x_0$.  The excess work along such a path is given by
\begin{equation*}
\begin{split}
W_\mathrm{ex} (\bar x_0 , - \nabla \Phi_1) 
\; & =\; - \int_0^{\infty}\!dt \, [\nabla U (x_t) +
q\, \nabla \Phi_1(x_t)] \cdot \, \dot x_t \\
\; & =\; (U + q \Phi_1) (\bar x_0) \;-\; (U + q \Phi_1) (\bar x_1).   
\end{split}
\end{equation*}
The right hand side of the previous equation represents the
quasi-potential corresponding to the final equilibrium state evaluated
at the initial state $\bar x_0$. 
More precisely, denote by $\bb P^\beta_{\bar x}$, $\bar x\in \bb R^d$,
the distribution of the process $X_t$ which solves \eqref{22} starting
from $\bar x$. One is interested in the asymptotic behavior of $X_t$
as $\beta \rightarrow \infty$. Fix a time interval $[T_-, T_+]$ and a
trajectory $x \colon [T_-, T_+] \to \bb R^d$. It is well known \cite{fw}
that
\begin{equation*}
\bb P^\beta_{\bar x} [X_t \approx  x_t \,,\, T_-\le t\le T_+ ] 
\asymp
\exp \Big\{ - \beta \, I_{[T_-, T_+]} (x|\bar x) \Big\},
\end{equation*}
where
\begin{equation*}
\label{1.3}
I_{[T_-, T_+]} (x|\bar x)=\frac 1{4\gamma} \int_{T_-}^{T_+}\!dt\,   
\big| \gamma \dot x_t + \nabla U(x_t) - q E(t,x_t) \big|^2 .
\end{equation*}
if $x(0)=\bar x$ and $I_{[T_-, T_+]} (x|\bar x)=+\infty$, otherwise.

Fix a time independent field $E$ and denote by $V_{E}\colon \bb R^d \to \bb
R_+$ the quasi potential defined by
\begin{equation*}
V_{E} (\bar x) \;=\; \inf I_{(-\infty, 0]} (x|\bar x_0),
\end{equation*}
where the infimum is carried over all paths $x(t)$ such that $x(0) =
\bar x$, $\lim_{t\to - \infty} x(t) = \bar x_0$ in which $\bar x_0$
belongs to the global attractor. When $E$ is gradient,
$E = - \nabla \Phi$, it is well known that $V_{\Phi} = U + q \Phi$
up to an additive constant that is fixed by requiring $V_{\Phi}
(\bar x_{\Phi})=0$.  In this case, the quasi potential
coincides with the excess work as computed above.
The corrections to the above result when a finite time window is
considered has been recently analyzed in \cite{AGMMM}.

Denote by $\mu_{\beta, E}$ the stationary distribution for the Langevin
dynamics \eqref{22}. When $E = - \nabla \Phi$, it is well known that the
distribution $\mu_{\beta, E}$ is proportional to $\exp\{ - \beta
(U+ q\Phi)\}$ and one can show that the relative entropy, as defined in
\eqref{relen}, satisfies
\begin{equation*}
\lim_{\beta\to\infty} \frac 1\beta S \big(\mu_{\beta, \Phi_0} \big| 
\mu_{\beta, \Phi_1} \big) \;=\; V_{\Phi_1} (\bar x_0).
\end{equation*}

\medskip
We turn to the nonequilibrium case, i.e. when $E$ is not a
gradient field. Assume that equation \eqref{23} has a unique global
attractor, e.g. an equilibrium point or a periodic orbit.  When $E$ is
not a gradient there are no simple expression for the quasi potential,
but it can be characterized as a solution of the stationary
Hamilton-Jacobi equation
\begin{equation*}
\big| \nabla V_E \big|^2 \;+\; 
\nabla V_E \cdot \big[ - \nabla U + q E \big] \;=\; 0.
\end{equation*}
Decompose the vector field $-\nabla U + q E$ as the sum of two orthogonal
pieces, $J_\mathrm{S}$ and $J_\mathrm{A}$, where
\begin{equation*}
J_\mathrm{S} \;=\; - \nabla V_E\;, \quad J_\mathrm{A} \;=\; -\nabla U
+ q E  +\nabla V_E \;, 
\end{equation*}
so that $-\nabla U + q E = J_\mathrm{S} + J_\mathrm{A}$. 
It follows from the Hamilton-Jacobi equation that $J_\mathrm{S}$, 
$J_\mathrm{A}$ are pointwise orthogonal, $J_\mathrm{S} (x) \cdot
J_\mathrm{A} (x) =0$ for all $x\in \bb R^d$.

Recall the expression \eqref{20} for the work done by the applied time
dependent field $E(t)$ in the time interval $[0, T]$ and define
the renormalized work, denoted by $W^\textrm{ren}_{[0, T]}$, as
\begin{equation}
\label{24}
\begin{split}
W^\textrm{ren}_{[0, T]}(\bar x_0, E) \; & =\; W_{[0, T]} (\bar x_0, E)
\;-\; \frac 1\gamma \int_{0}^{T}\! dt \, 
\big| J_\mathrm{A} (t,x(t)) \big|^2 \\
\;& =\; U(x(T)) - U(x(0))
\;+\; \frac 1\gamma \int_{0}^{T} \!dt  \, |J_\mathrm{S}(t,x(t)) |^2.
\end{split}
\end{equation}
where, as in \eqref{Weff}, we compute the quasi potential
$V_{E(t)}$ with $t$ frozen and denoted by $J_\mathrm{S}(t)$, 
$J_\mathrm{A}(t)$ the corresponding decomposition of the applied
field. 
The validity of the Clausius inequality for the renormalized work 
follows immediately from \eqref{24}. Moreover, by arguing as in
Section~\ref{s:4}, it is simple to check that equality is achieved in
the quasi static limit. 

Define the \emph{excess} work as the difference between the
renormalized work and the one involved in a quasi static
transformation:
\begin{equation*}
W_\mathrm{ex}(\bar x_0, E)  \;=\;
\frac 1\gamma \int_0^\infty \!dt \, |J_\mathrm{S}(t,x(t)) |^2.
\end{equation*}

Fix a point $\bar x$ and an external field $E_1$ constant in time.
Consider the path $\gamma \dot x_t \;=\; - \nabla U (x_t) \;+\; q\,
E_1(x_t)$ with initial condition $\bar x$.  Computing the excess
work along this path we get
\begin{equation*}
W_\mathrm{ex} (\bar x, E_1) = V_{E_1} (\bar x).
\end{equation*}

\medskip
To illustrate the previous definitions, consider the Langevin equation
\eqref{22} in two dimensions with $U(x) = (\lambda/4) |x|^4$ and
\begin{equation*}
E(t,x) \;=\; \alpha(t) \, x \;+\; A_0 \, \frac {x^\perp}{|x|}\;,
\end{equation*}
where $A_0>0$, $\lambda>0$, $\alpha (t)$ is a positive function, and
for $x=(x_1, x_2)$ we set $x^\perp = (-x_2, x_1)$.

Assume that $\alpha$ does not depend on $t$ and let $r_\alpha$ be the
minimum of $U(r) - (q/2) \alpha r^2$, $r_\alpha = \sqrt{q\alpha/
  \lambda}$. The deterministic flow defined by \eqref{23} has then 
the limit cycle $x(t) = r_\alpha (\cos (\omega t), \sin (\omega t))$, where
$\omega = A_0 q/\gamma r_\alpha$. 
The quasi potential is given by $V_\alpha (x) = U(x) - (q/2) \alpha
|x|^2 - (q^2\alpha^2/2\lambda)$ so that $J_\mathrm{S}(x) = -\nabla U
(x) + q\alpha x$, $J_\mathrm{A}(x) = q A_0 (x^\perp/|x|)$.  The power
dissipated along the periodic orbit is $\gamma r_\alpha^2 \omega^2$ so
the energy dissipated in an infinite time window is infinite.

Fix $\alpha_0 \not = \alpha_1$ and consider a function $\alpha (t)$
such that $\alpha (0) = \alpha_0$, $\alpha (t) = \alpha_1$, $t\ge
T$. Let $x(t)$ be the solution of \eqref{23} with initial condition
$\bar x$. The renormalized work and the excess work along such path
are given by
\begin{equation*}
\begin{split}
& W^\textrm{ren}(\bar x, E) \; =\; U(\bar x_1) - U(\bar x)
\;+\; \frac 1\gamma \int_{0}^{\infty} \!dt  \, | \nabla U (x(t)) -
q\alpha(t)  x(t)|^2, \\
& \qquad W_\mathrm{ex}(\bar x, E)  \;=\; \frac 1\gamma
\int_0^\infty \!dt \, | \nabla U (x(t)) - q\alpha(t)  x(t) |^2 \;,
\end{split}
\end{equation*}
where $\bar x_1$ is a point in the limit cycle corresponding to
$\alpha_1$. If the initial condition $\bar x$ belongs to the limit
cycle corresponding to $\alpha_0$, the previous integral is equal to
\begin{equation*}
\int_0^\infty\!dt\, \big( \partial_t  V_{\alpha (t)} \big)
(x(t)) \;=\; \frac {q^2}{4\lambda} (\alpha^2_1 - \alpha^2_0)
\;- \; \frac q2 \int_0^\infty\!dt\,   \dot \alpha(t) \,
|x(t)|^2 \;\ge\; 0.
\end{equation*}

What we have done is very close to the well known paper by Hatano and
Sasa \cite{hs}. The main difference is that we are considering the
limit of small noise in order to relate the quasi potential to the
work involved in the transformations. In particular, our
$W^{\text{ren}}$ is not a random variable.  There is also a difference
in terminology as we call $W^{\text{ren}}$ what they would call
$W^{\text{ex}}$, while we reserved this notation for a quantity which
is more closely related to the quasi potential.

\subsubsection*{Time dependent quasi potential}

To illustrate the time dependent quasi potential, consider the time
dependent Langevin equation with linear drift
\begin{equation*}
  \dot X_t \;=\; B(t) X_t  \;+\; E(t)  \;+\; 
  \sqrt{\frac 2\beta}\, \dot W_t
\end{equation*}
where $E(t)\in \mathbb R^n$ and $B(t)$ is a $n\times n$
time dependent matrix. As $X_t$ is a Gaussian process, its
distribution can be computed explicitly for any $\beta>0$. In
particular, the covariance and the mean of its distribution at time
$t$ can be recovered from the expression of the time dependent quasi
potential given below.

As $\beta \to \infty$ the evolution of $X_t$ in the time interval
$[T_1, T_2]$ satisfies a large deviations principle with rate function
\begin{equation*}  
  I_{[T_1,T_2]} (x)  \;=\; \frac 14 \int_{T_1}^{T_2} dt \,
  \big | \dot x(t)-B(t)\,x(t)- E (t)\big|^2 \,.
\end{equation*}
The associated time dependent Hamilton-Jacobi equation is
\begin{equation}
\label{26}
\partial_t V (t,x) \;+\; \big| \nabla V (t,x) \big|^2 
\;+\; \nabla V(t,x)\cdot [B(t)\,x + E(t)] \;=\; 0.
\end{equation}

Assume that $E(t)$, $B(t)$ are such that $(E(t), B(t))\to (E_0, B_0)$,
as $t\to - \infty$, $(E(t), B(t))\to (E_1, B_1)$ as $t\to\infty$,
respectively, and that the eigenvalues of $B_0$, $B_1$ have strictly
negative real part. Let $m_0 = - B_0^{-1} E_0$ and  $S_0$ be the
symmetric $n\times n$ matrix such that $S_0^2 = - (S_0B_0 +
B^T_0S_0)/2$. Let $S(t)$, $m(t)$ be the solution of
\begin{equation*}
    \left\{
    \begin{array}{l}
      \dot S=-2S^2-[SB+B^TS]\,,\\
      \dot m=Bm+E,
    \end{array}
  \right.
\end{equation*}
with boundary conditions $S(-\infty) = S_0$, $m(-\infty) = m_0$.
Then
\begin{equation*}
  V(t,x) \;=\; \frac 12 [x-m(t)] \cdot S(t) [x-m(t)]
\end{equation*}
is the solution of the time dependent Hamilton-Jacobi equation.  
As $t\to\infty$, $(S(t), m(t))$ converge to $(S_1,m_1)$, where $S_1^2
= - (S_1B_1 + B^T_1S_1)/2$ and $m_1= - B_1^{-1} E_1$. If $B_1$ is
normal, i.e.\ $B_1B_1^T=B_1^TB_1$, then  $S_1 = -(1/2) (B_1 + B^T_1)$.

In the one dimensional case, with $B=-(1/\theta)$, $\theta>0$, we get
\begin{equation*}
  m(t) \;=\; \int_{-\infty}^t ds\, \exp\Big\{ - \frac{t-s}\theta \Big\} 
  E(s)\;, \quad S(t) = \frac 1\theta\;\cdot
\end{equation*} 
In particular, $m(t) \to \theta E_0$, as $t\to-\infty$, $m(t) \to
\theta E_1$, as $t\to\infty$. When $\theta\ll 1$, that is when the
system relaxes very fast, the time dependent quasi potential at time
$t$ becomes the quasi potential computed with time frozen at $t$,
$V(t, \cdot) \approx V_{E(t)}(\cdot)$.

\subsection*{Acknowledgments}

We are grateful to J.\ Lebowitz for his insistence on a thermodynamic
characterization of the quasi potential. 
We acknowledge stimulating discussions with 
T.\ Komatsu, N.\ Nakagawa, S.\ Sasa, and H.\ Tasaki.
We also thank F.\ Flandoli for useful comments on hydrodynamic
equations with time dependent boundary conditions. 
We thank a referee for several comments and questions which have led
to an improvement of our paper.

\end{document}